\documentclass[reqno,11pt]{amsart}

\usepackage{hyperref}
\usepackage[mathscr]{eucal}
\usepackage{enumerate}
\numberwithin{equation}{section}

\newtheorem{Def}{Definition}[section]
\newtheorem{Thm}[Def]{Theorem}

\newtheorem{Lemma}[Def]{Lemma}

\newtheorem{Corollary}[Def]{Corollary}

\newcommand{\beq}{\begin{equation}}
\newcommand{\eeq}{\end{equation}}
\newcommand{\Proof}{\begin{proof}}
\newcommand{\QED}{\end{proof} \noindent}

\newcommand{\msp}{\hspace{-.1cm}}
\newcommand{\Riem}{{\rm Riem}}
\newcommand{\mm}{\hspace{-.08cm}\cdot \hspace{-.08cm}}

\newcommand{\R}{\mathbb{R}}

\newcommand{\Gammati}{\tilde{\Gamma}}

\allowdisplaybreaks[1]

\title[The RT-equations]{How to smooth a crinkled map of spacetime:\\ Uhlenbeck compactness for $L^\infty$ connections and optimal regularity for general relativistic shock waves by the Reintjes-Temple-equations}

\author[M.\ Reintjes]{Moritz Reintjes}
\address{Fachbereich f\"ur Mathematik und Statistik \\ Universit\"at Konstanz \\ D-78467 \\ Germany}
\email{moritzreintjes@gmail.com}
\thanks{M. Reintjes is currently supported by the German Research Foundation, DFG grant FR822/10-1, and was supported by FCT/Portugal through (GPSEinstein) PTDC/MAT-ANA/1275/2014 and UID/MAT/04459/2013 from January 2017 until December 2018.}

\author[B.\ Temple]{Blake Temple \\ \\ March 14, 2020 }
\address{Department of Mathematics\\ University of California\\ Davis, CA 95616\\ USA}
 \email{temple@math.ucdavis.edu}

\begin{document}

\begin{abstract}
We present authors' new theory of the RT-equations, nonlinear elliptic partial differential equations which determine the coordinate transformations which smooth connections $\Gamma$ to optimal regularity, one derivative smoother than the Riemann curvature tensor ${\rm Riem}(\Gamma)$. As one application we extend Uhlenbeck compactness from Riemannian to Lorentzian geometry; and as another application we establish that regularity singularities at GR shock waves can always be removed by coordinate transformation. This is based on establishing a general multi-dimensional existence theory for the RT-equations,  by application of elliptic regularity theory in $L^p$ spaces. The theory and results announced in this paper apply to arbitrary $L^\infty$ connections on the tangent bundle $T\mathcal{M}$ of arbitrary manifolds $\mathcal{M}$, including Lorentzian manifolds of General Relativity.    
\end{abstract}

\maketitle

\section{Introduction}

Although the Einstein equations of General Relativity (GR) are covariant, solutions are constructed in coordinate systems in which the PDE's take on a solvable form.  A very first question in GR is then, which properties of the spacetime represent the true geometry, and which are merely anomalies of the coordinate system?  In particular, does a solution of the Einstein equations exhibit its optimal regularity in the coordinate system in which it is constructed \cite{Israel,SmollerTemple}? At the low regularity of GR shock waves this goes to the question as to whether spacetime singularities are essential or removable.  Since coordinate systems define the local property of spacetime, the coordinates in which the metric is most regular determine the degree to which the physics in curved spacetime corresponds locally to the physics of Special Relativity.  It turns out that the problem of optimal regularity, the problem of constructing coordinates in which a connection is one derivative more regular than the curvature,\footnote{We say a connection (and its metric in the case of metric connections) exhibits {\it optimal regularity} in a given coordinate system if the connection is one (and the metric two) derivatives more regular than its Riemann curvature tensor, (c.f. Definition \ref{Def_regularity_singularity} below). This notion requires of course a choice of scale to measure the derivative, like $C^m$, $C^{m,\alpha}$ or $W^{m,p}$ \cite{Evans}. For convenience we here use the Sobolev space $W^{m,p}$ of functions with $m$ derivatives in $L^p$. Note, our main concern is the gain of one derivative because this suffices for Uhlenbeck compactness, not the level of $p$. By definition a metric is always one order more regular than its connection, and a connection is at most one derivative more regular than its Riemann curvature tensor.} is intimately related to Uhlenbeck compactness. 

In this paper we present the authors' new theory of the RT-equations, (the ``Regularity Transformation'' equations or ``Reintjes-Temple'' equations \cite{ReintjesTemple_ell1,ReintjesTemple_ell2,ReintjesTemple_RT_shocks}), a system of nonlinear elliptic partial differential equations which determines the local coordinate transformations which smooth a connection $\Gamma$ to optimal regularity, one derivative smoother than the Riemann curvature tensor ${\rm Riem}(\Gamma)$. The RT-equations apply to general connections $\Gamma$ (with or without torsion) on the tangent bundle $T\mathcal{M}$ of arbitrary $n$-dimensional manifolds $\mathcal{M}$, including Lorentzian manifolds of General Relativity.  The RT-equations are {\it elliptic} regardless of metric signature because they are constructed within the Euclidean Cartan algebra of differential forms associated with arbitrary coordinate systems, independent of any spacetime metric. A fully multi-dimensional existence theory for the RT-equations has now been established by application of elliptic regularity theory in $L^p$ spaces, and this is precisely what is needed to extend Uhlenbeck compactness from Riemannian to Lorentzian geometry.    By this, Uhlenbeck compactness and optimal regularity are pure logical consequences of nothing more or less than the rule which defines how connections transform from one coordinate system to another.

In this exposition of our theory, we derive the RT-equations, and present two existence theorems for the RT-equations together with the Uhlenbeck compactness theorems they imply:   (1)  If $\Gamma, {\rm Riem}(\Gamma) \in W^{m,p}$, then solutions to the RT-equations exist and furnish $W^{m+2,p}$ coordinate transformations which smooth $\Gamma$ to $W^{m+1,p}$ for $m\geq1$, $p>n$.   (2)  If $\Gamma, {\rm Riem}(\Gamma) \in L^{\infty}$, then for any $p>n$, solutions to the RT-equations exist and furnish coordinate transformations in $W^{2,2p}$ which smooth $\Gamma$ to $W^{1,p}$.\footnote{The space $W^{m,p}(\Omega)$ is the Banach space of functions $u \in L^p(\Omega)$ such that all weak derivatives up to order $m$ also lie in $L^p(\Omega)$, with norm $\|u\|_{W^{m,p}}= \sum_{|\beta|\leq m} \|D^\beta u\|_{L^p}$, where $\beta$ denotes the standard multi-index, c.f. \cite[Chapter 5.2]{Evans}.}  The latter demonstrates that regularity singularities do not exist at GR shock waves, and gives the first proof that metrics in weak solutions of the Einstein-Euler equations constructed by the Glimm scheme are one derivative more regular than previously known---regular enough to admit geodesics and locally inertial frames. The proofs of existence of solutions to the RT-equations within the above regularity classes is based on an iteration scheme which provides a numerical algorithm for smoothing connections to optimal regularity. To give the flavor of the proofs, we outline the argument in case (1), the case when existence and regularity follow from standard elliptic PDE theory without the need to modify and re-interpret the RT-equations, c.f. \cite{ReintjesTemple_ell2,ReintjesTemple_RT_shocks}.  In words, a crinkled map of spacetime, (i.e., one for which the metric is non-optimal), can always be smoothed by coordinate transformation, and we have an algorithm for doing it. 

Solutions of the Einstein equations which exhibit non-optimal metric regularity are well known and play an important role in the subject of GR shock waves, \cite{Israel,SmollerTemple,GroahTemple,ReintjesTemple1,VoglerTemple,LeFlochStewart}, and this was the starting point of authors' investigations. But the problem of the optimal regularity of connections is a much larger issue.  Non-optimal metrics and connections, with or without symmetries, exist as a direct consequence of Riemann's construction of a curvature tensor out of second derivatives of the metric. Indeed, if in a given coordinate system a metric exhibits optimal regularity in the sense that it is two derivatives more regular than its Riemann curvature,  then coordinate transformations with Jacobians one derivative less regular than the metric, will in general, by the tensor transformation law, lower the regularity of the metric by one order but preserve the regularity of the curvature, the latter because it is a tensor already one order less regular than the Jacobian. Similarly, since the transformation law for connections involves derivatives of the Jacobian, if we start with a connection $\Gamma$ of optimal regularity in the sense that its components are one derivative more regular than $Riem(\Gamma)$ in a given coordinate system, then coordinate transformations by Jacobians in the same regularity class as $\Gamma$ will transform $\Gamma$ to non-optimal regularity; i.e., the transformation will maintain the regularity of the components of $Riem(\Gamma)$, but will lower the regularity of the components of $\Gamma$ by one derivative, down to the regularity of $Riem(\Gamma)$.   The question we raise and answer here is whether this can always be reversed.   

That is, if a connection on the tangent bundle of a manifold  is non-optimal in the sense that it is less than one derivative more regular than its curvature in some given coordinate system, can it always be smoothed to optimal regularity by a ``singular''  coordinate transformation which removes irregularities in the highest order derivatives of the connection, but preserves the regularity of the curvature? Or is non-optimality a geometric property of connections on manifolds? Naively, it appears that not all connections could be smoothed to optimal regularity because the curvature does not involve all of the derivatives of the connection,  only the Curl. Nevertheless, Uhlenbeck proved that for connections on vector bundles over Riemannian manifolds (with positive definite metric) all derivatives are bounded by the curvature in Coulomb gauge. This is the basis of Uhlenbeck compactness, (c.f. \cite{Uhlenbeck}, topic of the 2019 Abel prize and 2007 Steele prize). Uhlenbeck compactness in \cite{Uhlenbeck} led to many important applications in Riemannian geometry, c.f. \cite{Taubes,Donaldson}. However, how to extend optimal regularity and Uhlenbeck compactness in general, from Riemannian geometry to the semi-Riemannian geometry of Physics, has long been an open problem.\footnote{The first optimal regularity result in Lorentzian geometry is due to Anderson \cite{Anderson}.  Anderson's results are based on using harmonic coordinates on the Riemannian hypersurfaces of a given foliation of spacetime, and establish curvature bounds for vacuum spacetimes and certain matter fields when the Riemann curvature is in $L^\infty$, under further assumptions. A similar result for vacuum spacetimes was proven in \cite{ChenLeFloch}.} By the RT-equations it is now clear that metric signature is of no relevance to the problem of optimal regularity and Uhlenbeck compactness---they follow just from the transformation law for connections.             

The RT-equations are defined in terms of the Cartan algebra of differential forms associated with a coordinate system $x$ in which the components $\Gamma^i_{jk}(x)$ of the connection $\Gamma$ are assumed to be given.  In this sense the RT-equations are not defined invariantly.  Recall that in differential geometry,  the exterior derivative $d$ is defined invariantly, independent of metric, but the co-derivative $\delta$ and Laplacian $\Delta=d\delta+\delta d$ must be defined in terms of an inner product induced on the Cartan algebra by some underlying metric, \cite{Dac}. Since the Riemann curvature tensor only involves derivatives of the connection through the metric-independent exterior derivative $d$,  it follows that if the components of the curvature and connection in a given coordinate system have the same regularity, then $d\Gamma$ has that regularity, and the co-derivative $\delta\Gamma$ (associated with the Euclidean coordinate metric) encodes the derivatives not directly controlled by the curvature (by Gaffney's Inequality \eqref{Gaffney}). Our idea, then, is that to get the classical Laplacian into the RT-equations, we introduce the Euclidean metric in $x$-coordinates as an auxiliary Riemannian structure, in place of, say, the invariant Lorentzian metric in the case of GR.    We then take $\delta$ to be the co-derivative of that Euclidean metric, which implies $\Delta=d\delta+\delta d$ is the classical (Euclidean) Laplacian.   By this, the RT-equations  are elliptic and the leading order operators are $d$, $\delta$ and $\Delta$ in every coordinate system. The right hand side of the RT-equations is nonlinearly coupled through operations defined on the Euclidean Cartan algebra, c.f. Section \ref{Sec_Prelim}.   

We now compare this to Uhlenbeck's method in  \cite{Uhlenbeck}.   Uhlenbeck's compactness theorem, Theorem 1.5 of \cite{Uhlenbeck},  applies to Riemannian metrics, and is based on establishing a uniform bound on the components of a connection in Coulomb gauge, the Coulomb gauge providing a coordinate system arranged to satisfy $\delta\Gamma=0$ to bound the derivatives uncontrolled by the curvature.  This works when $\delta$ is taken to be the co-derivative of the invariant Cartan algebra of the underlying Riemannian metric of the connection $\Gamma$.   Uhlenbeck compactness in Coulomb gauge then follows from a uniform bound on the curvature.  To illustrate the heart of the issue in \cite{Uhlenbeck}, taking $\delta$ of ${\rm Riem}(\Gamma)=d\Gamma+\Gamma\wedge\Gamma$, when $\delta\Gamma=0$, results in an equation of (essentially) the form $\Delta\Gamma=\delta\, {\rm Riem}(\Gamma)$, where $\Delta=d\delta+\delta d$ is the Laplace-Beltrami operator of the underlying (positive definite) Riemannian metric up to lower order corrections; so by elliptic regularity, a sequence of connections $\Gamma_i$ with ${\rm Riem}(\Gamma_i)$ uniformly bounded in $L^p$, will be uniformly bounded in $W^{1,p}$ in Coulomb gauge, for $p<\infty$. Sobolev compactness then implies a subsequence converges in $L^p$ in Coulomb gauge.  The important difference is that the Coulomb gauge condition $\delta\Gamma=0$ requires that $\delta$ and $\Gamma$ both be associated with the {\it same} Riemannian metric.   In the Lorentzian case, Uhlenbeck's argument leads to a hyperbolic equation in Coulomb gauge, and obtaining optimal regularity from a hyperbolic equation is problematic, (c.f. discussion in Section \ref{Sec_Discussion}).   Alternatively, the RT-equations are constructed from the co-derivative $\delta$ of the Cartan algebra induced by an auxiliary Euclidean metric, so in general one cannot expect a gauge in which $\delta\Gamma=0$. Our result can be viewed as establishing the existence of a coordinate transformation under which $\delta\Gamma$ has the same regularity as $d\Gamma$, so  Gaffney's Inequality implies optimal regularity without requiring the more constraining Coulomb condition $\delta{\Gamma}=0$. Our analysis does not assume optimal regularity of the connection at the start, (c.f. Theorem 1.3 of \cite{Uhlenbeck} which assumes $\Gamma\in W^{1,p}$ without estimate), but establishes optimal regularity assuming the connection is estimated in the same space as the curvature.  Uhlenbeck's theorem applies to connections on vector bundles over Riemannian base manifolds, and our theorem extends this to arbitrary base manifolds. Here we address the case of the tangent bundles and we currently work on extending this to general vector bundles over arbitrary base manifolds.   

Our theorems on regularizing non-optimal solutions of the Einstein equations in GR provide a completely general four dimensional theory, but as an application they also establish for the first time that non-optimal solutions of the Einstein equations constructed by the Glimm scheme in spacetimes with symmetry, are one order more regular than previously known.   Shock waves are the lowest regularity solutions of the Einstein equations which incorporate realistic perfect fluid sources. Shock waves are fundamental because they introduce time irreversibility, increase of entropy and dissipation into the evolution of GR perfect fluids without violating causality, and without the need to introduce a consistent relativistic theory of viscosity and  heat conduction, \cite{FreistuehlerTemple}. The first existence theory by the Glimm scheme was given in \cite{GroahTemple}, and this could only be accomplished in Standard Schwarzschild Coordinates (SSC), coordinates in which the metric is always non-optimal as a consequence of the link between the radial coordinate and the area of the spheres of symmetry.  \footnote{The fact that there exist non-optimal coordinate systems in which the Einstein equations are simple enough to implement a Glimm type analysis is serendipitous, but we emphasize that non-optimal coordinates play no special role in spherical symmetry--they exist simply because the curvature involves second derivatives of the metric, but transforms as a tensor. See also \cite{LeFlochStewart} for non-optimal solutions in spacetimes with different symmetries.}    Thus the gravitational metric is only Lipschitz continuous ($C^{0,1}$) at shock waves in SSC, even though both the connection and curvature tensor of such solutions stay bounded in $L^{\infty}$. It has been an open question as to whether such $C^{0,1}$ metrics can always be smoothed one order to optimal metric regularity by coordinate transformation. This question is deeply related to the existence and regularity of locally inertial coordinate systems, and thus to the local correspondence of GR with the physics of Special Relativity. For example, H\"older continuity for the connection is required to prove existence of geodesics by Peano's Theorem, and optimal connection regularity $W^{1,p}$ for $p>n$ gives H\"older continuity at shocks by Morrey's inequality. In the RSPA publication \cite{ReintjesTemple1} authors conjectured that if such coordinate systems do not exist, then shock wave interactions create a new kind of singularity in GR which the authors termed {\it regularity singularity},  (see also \cite{Reintjes,ReintjesTemple2}), a point in spacetime at which the metric connection fails to have optimal regularity in any coordinate system. The results in this paper imply that no regularity singularities exist in General Relativity for $L^\infty$ curvature. 

The question as to the existence of such smoothing transformations is surprisingly subtle. The Riemann normal construction is not sufficient to smooth a metric and its connection to optimal regularity, and the construction itself is problematic for Lipschitz ($C^{0,1}$) metrics.    At smooth, non-interacting shock surfaces, a now classical result of Israel shows that transformation to Gaussian normal coordinates at the surface suffices to smooth a $C^{0,1}$ metric to $C^{1,1}$ at the shock when the connection and the curvature are discontinuous but bounded across the shock surface \cite{Israel}. But for more general shock wave interactions, the only result we have since Israel is due to Reintjes \cite{Reintjes}, who proved that the gravitational metric can always be smoothed one order to $C^{1,1}$ in a neighborhood of the interaction of two shock waves from different characteristic families, in spherically symmetric spacetimes. Reintjes' procedure for finding the local coordinate systems of optimal regularity is orders of magnitude more complicated than the Riemann normal, or Gaussian normal construction process. The coordinate systems of optimal $C^{1,1}$ regularity are constructed in \cite{Reintjes} by solving a non-local PDE highly tuned to the structure of the interaction. Trying to guess the coordinate system of optimal smoothness apriori, for example harmonic or Gaussian normal coordinates \cite{Choquet}, didn't work.  In Reintjes' construction, several apparent miracles happen in which the Rankine-Hugoniot jump conditions come in to make seemingly over-determined equations consistent, but at this stage, the principle behind what PDE's must be solved to smooth the metric in general, or when this is possible, appears entirely mysterious.  

The authors' current point of view on the question of regularity singularities began with the formulation of the {\it Riemann-flat condition} in \cite{ReintjesTemple_geo}, a necessary and sufficient condition for the existence of a coordinate transformation which smooths a connection in $L^\infty$ to $C^{0,1}$. The Riemann-flat condition is the condition that there should exist a symmetric $(1,2)$-tensor $\Gammati$, one order smoother than the connection $\Gamma$, such that  ${\rm Riem}(\Gamma-\Gammati)=0$, remarkable because it is a geometric condition on $\Gammati$ alone, independent of the coordinate transformation that smooths the metric. Since $\Gamma$ and $\Gamma -\Gammati$ have the same singular set (shock set), at first we thought the Riemann-flat condition was telling us that to smooth an $L^\infty$ shock wave connection one needed to extend the singular shock set to a flat connection by some sort of Nash embedding theorem. Our point of view changed again with the successful idea that we might derive a system of elliptic equations equivalent to the Riemann-flat condition, which resulted in the Regularity Transformation equations, equations \eqref{eqn1} - \eqref{eqn4} below. 

Finally we comment on the issue of geometric invariance. The ellipticity of the RT-equations is an invariant property in the sense that a different version of them is expressed in terms of the Euclidean Laplacian in every coordinate system. They are not tensorial, but from an analytic point of view this is a virtue. The fact that the theory of the RT-equations employs no special coordinate gauge means the resulting Uhlenbeck compactness theorem applies to approximate solutions constructed in any coordinate system, making it inherently useful for analysis, which is virtually always accomplished in coordinate systems, not invariantly. By generality of the theorem, the approximating connections need not even be metric. 
To compare, the Einstein equations transform between coordinate systems in such a way that they appear simpler in certain canonical gauges--the RT-equations are equally simple in all gauges.

To place the RT-equations into the context of ``geometry'', note that if the atlas of a manifold is sufficiently regular, then the regularity of the metric, connection and curvature are the same in every coordinate system, (making regularity ``geometric''); but if the regularity of the atlas is suitably low to allow the possibility of lifting or lowering the regularity of the metric and connection while maintaining the regularity of the curvature, then ``regularity'', as well as ``optimal regularity'', are coordinate dependent notions.  Thus one should view the RT-equations as providing a low regularity coordinate transformation which lifts the regularity of non-optimal connection components to optimal regularity, one order of derivative above the curvature, but once lifted, restricting to  the smooth atlas then makes regularity, and optimal regularity, a coordinate independent ``geometric'' property. 
This dependence of regularity and optimal regularity on the atlas and coordinate system holds at every level of derivative, in every regularity class, and the only difference between one regularity and another (say $W^{1,p}$ vs $L^\infty$) is the issue of solving the RT-equations.  

In this paper we summarize the proofs in the case $\Gamma, d\Gamma\in W^{m,p}$, $m\geq 1$, $p>n$, the lowest regularity in which the RT-equations can be introduced, and existence and regularity established by standard elliptic regularity theory in $L^p$ spaces without the need to modify and re-interpret the equations. We refer the reader to \cite{ReintjesTemple_RT_shocks} for our extension of the $W^{1,p}$ existence theory to the case $\Gamma, d\Gamma \in L^{\infty}$ based on what we call the {\it reduced} RT-equations,  a modified version of the RT-equations amenable to analysis in $L^\infty$. We do not discuss the reduced RT-equations here, but state our results for $L^\infty$ connections in \cite{ReintjesTemple_RT_shocks} as Theorems \ref{Thm_opt_reg_shocks} and \ref{Thm_compactness_shocks} in the end of Section \ref{Sec_Results}.

In Section \ref{Sec_Results} we state our main theorems in terms of the Euclidean Cartan Algebra of differential forms summarized in Section \ref{Sec_Prelim}. Applications of our Uhlenbeck compactness theorem and of our optimal regularity result to General Relativity are presented in Section \ref{Sec_Applications}. The Riemann-flat condition is introduced in Section \ref{Sec_Riemann-flat}, and is the starting point for deriving the RT-equations in Section \ref{Sec_Proof_Thm1}. The existence theory for the RT-equations is presented in Section \ref{Sec_Proof_Thm2}. The uniform bounds on connections of optimal regularity, from which Uhlenbeck compactness is deduced, are established in Section \ref{Sec_bounds}. Section \ref{Sec_Discussion} contains a discussion of non-optimal solutions in the context of the initial value problem in GR.

\section{Statement of results}    \label{Sec_Results}

Our first theorem establishes the equivalence of the Riemann-flat condition with the solvability of the RT-equations.  Our second theorem gives an existence theorem for the RT-equations in the case $\Gamma$, $d\Gamma\in W^{m,p}$, (and hence ${\rm Riem}(\Gamma)\in W^{m,p}$), for $m\geq1$, $p>n$.   Combining the two theorems, we conclude in Theorem \ref{ThmMainCor} that any such connection can be mapped locally to optimal regularity by a coordinate transformation, demonstrating that regularity singularities do not exist when $\Gamma$ and $d\Gamma$ are in $W^{m,p}$, $m\geq1$, $p>n$. This is case (1) of the Introduction.  The gain of one derivative then suffices to conclude with Uhlenbeck compactness in Theorem \ref{Thm_compactness} below for the case $m\geq1$, $p>n$.  In Theorems \ref{Thm_opt_reg_shocks} and \ref{Thm_compactness_shocks}  we state our optimal regularity and Uhlenbeck compactness results for the case of $L^\infty$ connections, the low regularity class of GR shock waves. This is case (2) of the Introduction. By $\Gamma$, $d\Gamma\in W^{m,p}$ we mean the component functions of $\Gamma$ and $d\Gamma$ are in $W^{m,p}$ in some given, but otherwise arbitrary, coordinate system $x$, ($m\geq 0$ and $p>n$, including $p=\infty$).

We begin by giving our definition of optimal regularity. Because the problem of optimal regularity is local, without loss of generality, we restrict to a single coordinate chart $x$ defined on some open set in $\mathcal{M}$ with image $\Omega \subset \R^n$ open and  bounded with smooth boundary.   

\begin{Def}   \label{Def_regularity_singularity}
Let $\Gamma\equiv\Gamma^\mu_{\nu \rho}$ denote the components in $x$-coordinates of a connection defined on the tangent bundle $T\mathcal{M}$ of a manifold $\mathcal{M}$. Assume further that each component of its Riemann curvature tensor $Riem(\Gamma)$ is in $W^{m,p}(\Omega)$ for some $m\geq 0$, $p\geq 1$, but is no smoother in the sense that $Riem(\Gamma)$ is not in $W^{m',p}(\Omega)$ for any $m'>m$. We say $\Gamma$ has {\rm optimal regularity} in $x$-coordinates if $\Gamma \in W^{m+1,p}(\Omega)$, one order smoother than $Riem(\Gamma)$. 
\end{Def}

\noindent Again, our main concern is not the level of $p$, but the gain of one derivative from $m$ to $m+1$ because this suffices for Uhlenbeck compactness.

To state the first theorem, view $\Gamma\equiv\Gamma^\mu_{\nu k}dx^k$ as a matrix valued $1$-form defined in $x$-coordinates. The RT-equations depend on $\Gamma$ and the coordinate system $x$ in which the components of $\Gamma$ are given. The unknowns in the RT-equations are $\Gammati,J,A$,  taken to be matrix valued differential forms defined on the coordinate system $x$ as follows: Let $J\equiv J^\mu_\nu$ denote the Jacobian of the sought after coordinate transformation which smooths the connection, viewed as a matrix-valued $0$-form; let $\Gammati\equiv \Gammati^\mu_{\nu k}dx^k$ be an unknown tensor one order smoother than $\Gamma$ (as required for the Riemann-flat condition $Riem(\Gamma-\Gammati)=0$) viewed as a matrix-valued $1$-form; and let $A\equiv A^\mu_\nu$ be an auxiliary matrix valued $0$-form introduced to impose $Curl( J)=0$, the integrability condition for the Jacobian. 

\begin{Thm} \label{Thm_main}
Assume $\Gamma \equiv\Gamma^\mu_{\nu k}$ are the components in $x$-coordinates of a connection on $T\mathcal{M}$ of manifold $\mathcal{M}$, defined on $\Omega\subset{\mathbb R}^n$ open and bounded with smooth boundary.   Assume that $\Gamma\in W^{m,p}(\Omega)$ and $d\Gamma\in W^{m,p}(\Omega)$ for $m\geq 1,$ $p>n$, $p<\infty$. Then the following equivalence holds: \vspace{.15cm} \newline 
Assume there exists $J \in W^{m+1,p}(\Omega)$ invertible, $\Gammati \in W^{m+1,p}(\Omega)$  and $A\in W^{m,p}(\Omega)$ which solve the elliptic system
\begin{eqnarray} 
\Delta \Gammati &=& \delta d \Gamma - \delta \big( d(J^{-1})\wedge dJ\big) + d(J^{-1} A ), \label{eqn1} \\
\Delta J &=& \delta ( J \mm \Gamma ) - \langle d J ; \tilde{\Gamma}\rangle - A , \label{eqn2} \\
d \vec{A} &=& \overrightarrow{\text{div}} \big(dJ \wedge \Gamma\big) + \overrightarrow{\text{div}} \big( J\, d\Gamma\big) - d\big(\overrightarrow{\langle d J ; \tilde{\Gamma}\rangle }\big),   \label{eqn3}\\
\delta \vec{A} &=& v,  \label{eqn4}
\end{eqnarray}
with boundary data 
\begin{eqnarray}   
Curl(J) \equiv \partial_j J^\mu_i - \partial_i J^\mu_j =0 \ \ \text{on} \ \partial \Omega,  \label{bdd1}
\end{eqnarray}
where $v\in W^{m-1,p}(\Omega)$ is some vector valued $0$-form free to be chosen.
Then for each $q\in\Omega$,  there exists a neighborhood $\Omega'\subset\Omega$ of $q$ such that $J$ is the Jacobian of a coordinate transformation $x \to y$ on $\Omega'$, and the components of $\Gamma$ in $y$-coordinates are in $W^{m+1,p}(\Omega')$. \vspace{.15cm} \newline 
Conversely, if there exists a coordinate transformation $x \to y$ with Jacobian $J = \frac{\partial y}{\partial x} \in W^{m+1,p}(\Omega)$ such that the components of $\Gamma$ in $y$-coordinates are in $W^{m+1,p}(\Omega)$, then there exists $\Gammati \in W^{m+1,p}(\Omega)$ and $A\in W^{m,p}(\Omega)$ such that $(J,\Gammati,A)$ solve \eqref{eqn1} - \eqref{bdd1} in $\Omega$ for some $v\in W^{m-1,p}(\Omega)$.
\end{Thm}

Equations (\ref{eqn1})-(\ref{eqn4}) are the \emph{RT-equations}, a system of elliptic equations associated with each coordinate system $x$ which we consider fixed from now on. To derive the RT-equations we develop an Euclidean Cartan algebra of differential forms associated with coordinate system $x$, summarized in Section \ref{Sec_Prelim}. The starting point for this Cartan calculus is to take the Euclidean metric in $x$-coordinates as an auxiliary metric in terms of which we introduce the Euclidean Laplacian $\Delta \equiv d\delta + \delta d$ and the Euclidean co-derivative $\delta$, and because of this, the RT-equations are elliptic. Here $\vec{A}$, the vectorization of $A$, is the vector valued $1$-form defined by $\vec{A} \equiv A^{\mu}_{i}dx^i$, so $d\vec{A} = Curl (A)$. The operations $\vec{\cdot},$ $\overrightarrow{\rm div}$ and $\langle \cdot \; ; \cdot \rangle$ are defined in terms of the Cartan algebra as well. Equation \eqref{eqn3} is obtained by setting $d$ of the vectorized right hand side of (\ref{eqn2}) equal to zero, thus the identity $d\vec{J} = Curl (J)$ implies that \eqref{eqn3}  is equivalent to the integrability condition $Curl( J)=0$ for the Jacobian, c.f. \cite{ReintjesTemple_ell1}. The first two terms on the right hand side of \eqref{eqn3} result from identity \eqref{regularity-miracle} proven below. It is by this identity that seemingly uncontrolled terms involving $\delta\Gamma$  can be re-expressed in terms of the more regular $d\Gamma$, resulting in a fortuitous gain of one derivative required for the whole theory to work.

The derivation of the RT-equations in Section \ref{Sec_Proof_Thm1} shows that if $\Gammati$ satisfies the Riemann-flat condition, there exists $J$ and $A$ such that $(J,\Gammati,A)$ solve the RT-equations with the regularities required in Theorem \ref{Thm_main}. The converse is more subtle. The following lemma is the main step in the proof of the converse, i.e. the forward implication of Theorem \ref{Thm_main}, that existence for the RT-equations implies existence of local coordinate transformations which smooth the connection $\Gamma$ to optimal regularity:      

\begin{Lemma}  \label{Lemma_main}
Let $\Gamma, d\Gamma\in W^{m,p}(\Omega)$  be given, for $m\geq 1,$ $p>n$, $p<\infty$. Assume $(J,\Gammati,A)$ solves the RT-equations \eqref{eqn1} - \eqref{eqn4}, then 
\beq \label{modification} 
\Gammati' \equiv  \Gamma  - J^{-1} dJ 
\eeq 
solves the Riemann-flat condition ${\rm Riem}(\Gamma-\Gammati')=0$, and $\Gammati'$ has the regularity of $\Gammati$.
\end{Lemma}

\noindent To explain, the solution $(\Gammati,J,A)$ of the RT-equations has the correct regularity and produce  the correct Jacobian $J$, but not the correct $\Gammati$ because $\Gammati$ might not satisfy the Riemann-flat condition. The miracle of Lemma \ref{Lemma_main} is that $\Gammati'$, (which solves the Riemann-flat condition), is defined in terms of $\Gamma$ and $J$ alone, but also solves the RT-equations for the same $J$ and thereby inherits the optimal regularity of $\Gammati$ due to cancellations between $J^{-1}dJ$ and $\Gamma$. The mapping from $\Gammati$ to $\Gammati'$ is a gauge transformation in the sense that $\Gammati'$ again solves the RT-equations, but for different matrix valued $0$-form $A'$ in place of $A$. The gauge freedom, i.e., the freedom to assign $v$, is a propitious feature of the RT-equations.    In particular, equation (\ref{eqn1}) was obtained by augmenting the first order Riemann-flat condition to a first order Cauchy-Riemann type system, and then extending the solution space by replacing this with the implied second order Poisson type equation (\ref{eqn1}), without imposing the nonlinear boundary data required to recover solutions of the original Cauchy-Riemann equations, c.f. Section 3.1 in \cite{ReintjesTemple_ell1}. By Lemma \ref{Lemma_main} we recover the Riemann-flat condition from the second order system by use of the gauge freedom in the equations, without having to solve the original first order system.

To state our theorems regarding existence, optimal regularity and Uhlenbeck compactness, we introduce the shorthand notation
\beq \label{norm_Gamma_dGamma}
\|(\Gamma,d\Gamma)\|_{W^{m,p}(\Omega)} 
\equiv \|\Gamma\|_{W^{m,p}(\Omega)} \: + \: \|d\Gamma\|_{W^{m,p}(\Omega)}.
\eeq
The following theorem states the first existence result for the RT-equations.

\begin{Thm}  \label{ThmMain}
Assume the components of $\Gamma, d\Gamma \in W^{m,p}(\Omega)$ for $m\geq 1$, $p>n\geq2$, $p<\infty$, in some coordinate system $x$. Then for each $q\in\Omega$ there exists a solution $(\Gammati,J,A)$ of the RT-equations \eqref{eqn1} - \eqref{eqn4} with boundary data \eqref{bdd1} defined in a neighborhood $\Omega_q$ of $q$ such that $\Gammati\in W^{m+1,p}(\Omega_q),$  $J\in W^{m+1,p}(\Omega_q)$ invertible and $A\in W^{m,p}(\Omega_q)$. Moreover, if $M\geq 0$ is a constant such that $\|(\Gamma,d\Gamma)\|_{W^{m,p}(\Omega)} \leq M$, then there exists a constant $C(M)>0$ such that 
\small
\beq   \label{bound_u_a_Thm}
\|I-J\|_{W^{m+1,p}(\Omega_q)} + \|\Gammati\|_{W^{m+1,p}(\Omega_q)} + \|A\|_{W^{m,p}(\Omega_q)} 
\leq \ C(M)\, \|(\Gamma,d\Gamma)\|_{W^{m,p}(\Omega_q)} ,
\eeq
\normalsize    
where $I$ is the identity, and both $\Omega_q$ and $C(M)$ depend only on $M, \Omega, m, n, p$.\footnote{Sobolev norms $\|\cdot\|_{W^{m,p}}$ are understood to be taken on all components of matrix or vector valued differential forms, and then summed.}
\end{Thm}

As an immediate corollary of Theorems \ref{Thm_main} and \ref{ThmMain} we deduce the optimal regularity theorem in case (1) of the Introduction.

\begin{Thm}  \label{ThmMainCor}
Assume the components of $\Gamma, d\Gamma \in W^{m,p}(\Omega)$ for $m\geq 1$, $p>n\geq2$, in $x$-coordinates, such that $\|(\Gamma,d\Gamma)\|_{W^{m,p}(\Omega)} \leq M$ for some constant $M\geq 0$. Then for each $q\in\Omega$ there exists a coordinate transformation $x \to y$ defined in a neighborhood $\Omega_q$ of $q$ (depending only on $M,\Omega, m, n, p$), with Jacobian $J \in W^{m+1,p}(\Omega)$ , such that the components of $\Gamma$ in $y$-coordinates, $\Gamma_y \equiv \Gamma^\gamma_{\alpha\beta}(y)$, are in $W^{m+1,p}(\Omega'_q)$ for any $\Omega_q'$ compactly contained in $\Omega_q$, and $\Gamma_y$ satisfies the uniform bound
\beq \label{curvature_estimate_Main}
\|\Gamma_y\|_{W^{m+1,p}(\Omega'_q)} \leq C(M) \|(\Gamma,d\Gamma)\|_{W^{m,p}(\Omega_q)},
\eeq
where $C(M)>0$ is some constant depending only on $\Omega_q, \Omega'_q, M, m, n, p$.
\end{Thm}

Estimate \eqref{curvature_estimate_Main} extends Uhlenbeck's curvature estimate, Theorem 1.3 (ii) of \cite{Uhlenbeck}, to connections on tangent bundles of arbitrary manifolds, including semi-Riemannian and Lorentzian manifolds, above the threshold $m\geq 1$, $p>n$. We outline the proof of Theorem \ref{ThmMain} below, and refer to \cite{ReintjesTemple_ell2} for further details. For the proof we introduce an iteration scheme designed to apply the linear theory of elliptic regularity in $L^p$ spaces. A key insight for the proof was to augment the RT-equations by ancillary elliptic equations in order to convert the non-standard boundary condition $Curl(J)=0$ in \eqref{bdd1}, which is of neither Neumann nor Dirichlet type, into Dirichlet data for $J$ at each stage of the iteration, c.f. \cite{ReintjesTemple_ell2}. By this, the iteration scheme can be defined and bounds sufficient to imply convergence in the requisite spaces can be proven,  by applying standard existence theorems regarding elliptic regularity in $L^p$ spaces for the linear Poisson equation, to each iterate, \cite{GilbargTrudinger}. The regularity $\Gamma, Riem(\Gamma)\in W^{m,p}$, $m\geq1 ,p>n,$ is a natural threshold, because this is the lowest regularity that implies $\Gamma, Riem(\Gamma)$ are H\"older continuous by Morrey's inequality, c.f. \eqref{Morrey_textbook} below. This is used in the proof of convergence of the iteration scheme to control the nonlinear products on the right hand side of the RT-equations by point-wise estimates.
We now state the Uhlenbeck compactness theorem for case (1) of the Introduction, which results from the uniform bound \eqref{curvature_estimate_Main}.

\begin{Thm}   \label{Thm_compactness}
Assume $\Gamma_i \equiv (\Gamma_i)^\sigma_{\mu\nu}$ are the $x$-components of the $i$-th connection in a sequence of connections $\{\Gamma_i\}_{i\in \mathbb{N}}$ on the tangent bundle $T\mathcal{M}$ of an $n$-dimensional manifold $\mathcal{M}$, all given in a fixed coordinate system $x$ on $\Omega$. Let $m\geq 1$,  $p \in (n,\infty)$ and $n\geq 2$. Assume that in $x$-coordinates $\Gamma_i, d\Gamma_i \in W^{m,p}(\Omega)$ satisfy the uniform bound
\beq \label{uniform_bound2}
\|(\Gamma_i,d\Gamma_i)\|_{W^{m,p}(\Omega)}  \leq M
\eeq
for some constant $M>0$ independent of $i\in \mathbb{N}$. Then for every $q \in \Omega$ there exists a neighborhood $\Omega' \subset \Omega$ of $q$, and a subsequence of $\Gamma_i$, (also denoted by $\Gamma_i$), for which the following holds:

\noindent{\bf (i)} There exists for each $\Gamma_i$ a coordinate transformation $x \to y_i(x)$ defined on $\Omega'$, such that the components $\Gamma_{y_i} \equiv (\Gamma_i)_{y_i}$ of $\Gamma_i$ in $y_i$-coordinates exhibit optimal regularity, $\Gamma_{y_i}\in W^{m+1,p}(\Omega')$, with uniform bounds $W^{m+1,p}(\Omega')$ of the form \eqref{curvature_estimate_Main}. 

\noindent{\bf (ii)} The $y_i$-components $\Gamma_{y_i}$, taken as functions of $x$, also exhibit optimal regularity $\Gamma_{y_i}(x) \equiv (\Gamma_i)_{y_i}(y_i(x)) \in W^{m+1,p}(\Omega')$, with uniform bounds \eqref{curvature_estimate_Main} in $W^{m+1,p}(\Omega')$. 

\noindent{\bf (iii)} The transformations $x\to y_i(x)$ are uniformly bounded in $W^{m+2,p}(\Omega')$, and converge to a transformation $x\to y(x)$, weakly in $W^{m+2,p}(\Omega')$, strongly in $W^{m+1,p}(\Omega')$. 

\noindent{\bf (iv)} {\rm Main Conclusion:}   There is a subsequence of $\Gamma_{y_i}$ which converge to some $\Gamma_y(x)$, weakly in $W^{m+1,p}(\Omega')$, strongly in $W^{m,p}(\Omega')$. Moreover, letting $\Gamma_x$ denote the weak limit of the non-optimal $\Gamma_i$ in $W^{m,p}(\Omega')$ in $x$-coordinates, then the limit $\Gamma_y$ is the connection $\Gamma_x$ transformed to $y$-coordinates.
\end{Thm}

Theorem \ref{Thm_compactness} is a consequence of the uniform bound \eqref{curvature_estimate_Main} and application of the Banach Alaoglou compactness Theorem in $W^{m+1,p}$, together with the uniform bound \eqref{bound_u_a_Thm}, required to conclude the existence of a convergent subsequence of $y_i$, see \cite{ReintjesTemple_RT_shocks} for a careful proof. We now state our otpimal regularity and Uhlenbeck compactness theorems for case (2) of the Introduction, the case $\Gamma, d\Gamma \in L^\infty(\Omega)$, the low regularity associated with GR shock waves \cite{ReintjesTemple_RT_shocks}.

\begin{Thm} \label{Thm_opt_reg_shocks} 
Assume $\Gamma, d\Gamma \in L^\infty(\Omega)$ in $x$-coordinates, and let $M\geq 0$ be a constant with $\|(\Gamma,d\Gamma)\|_{L^\infty(\Omega)} \leq M$. Then, for any $p < \infty$ and for each $q\in\Omega$, there exists a coordinate transformation $x \to y$ defined in a neighborhood $\Omega_q$ of $q$ (depending only on $M,\Omega, n, p$), with Jacobian $J \in W^{1,2p}(\Omega_q)$ such that the components of $\Gamma$ in $y$-coordinates, $\Gamma_y \equiv \Gamma^\gamma_{\alpha\beta}(y)$, are in $W^{1,p}(\Omega'_q)$ for any $\Omega_q'$ compactly contained in $\Omega_q$, and $\Gamma_y$ satisfies the uniform bound
\beq \label{curvature_estimate_shockcase}
\|\Gamma_y\|_{W^{1,p}(\Omega'_q)} \leq C(M)\, \|(\Gamma,d\Gamma)\|_{L^\infty(\Omega_q)},
\eeq
where $C(M)>0$ is some constant depending only on $\Omega_q, \Omega'_q, M, m, n, p$.
\end{Thm}

As for Theorem \ref{Thm_compactness}, estimate \eqref{curvature_estimate_shockcase} on $\Gamma_y$ provides the uniform bound required to conclude with compactness by applying the Banach Alaoglu Theorem in $W^{1,p}$, which gives the following:  

\begin{Thm}\label{Thm_compactness_shocks}   
Assume $\{\Gamma_i\}_{i\in \mathbb{N}}$ is a sequence of connection components in $x$-coordinates such that $\|(\Gamma_i,d\Gamma_i)\|_{L^\infty(\Omega)} \leq M$. Then, for any $p \in (n,\infty)$, points (i) - (iv) of Theorem \ref{Thm_compactness} hold for $m=0$, but with $W^{2,2p}$ convergence of a subsequence of $\{y_i\}_{i\in \mathbb{N}}$.
\end{Thm}

Theorems \ref{Thm_compactness} and \ref{Thm_compactness_shocks} extend Uhlenbeck's compactness theorem for connections on vector bundles over Riemannian manifolds \cite{Uhlenbeck} to connections on tangent bundles of arbitrary (differentiable) manifolds, including Lorentzian manifolds of relativistic Physics. Note that Uhlenbeck's theorem in \cite{Uhlenbeck} assumes a uniform bound on the curvatures ${\rm Riem}(\Gamma_i)$ together with the assumption that $\Gamma_i$ is in the space of optimal regularity one derivative smoother than ${\rm Riem}(\Gamma_i)$ without an estimate. Our compactness results assume a uniform bound on ${\rm Riem}(\Gamma_i)$, does not assume that $\Gamma_i$ have optimal regularity, but requires that $\Gamma_i$ be uniformly bounded in the same space as the curvature ${\rm Riem}(\Gamma_i)$. See \cite{ReintjesTemple_RT_shocks} for a discussion of this point. 

Finally note, assuming $\Gamma, \Riem(\Gamma) \in L^\infty$,  Theorem \ref{Thm_opt_reg_shocks} establishes optimal regularity $\Gamma\in W^{1,p}$, which suffices to establish Uhlenbeck compactness and to construct geodesic curves and locally inertial frames for GR shock waves and General Relativity \cite{ReintjesTemple_RT_shocks}. However, it is still  open as to whether such $L^\infty$ connections can be smoothed to Lipschitz continuity ($W^{1,\infty} \equiv C^{0,1}$), since $p=\infty$ is a singular case of elliptic regularity theory. That is, there exist solutions of the Poisson equation which fail to be two derivatives more regular than their $L^\infty$ sources, c.f. \cite{ReintjesTemple_ell1}.

For a careful proof of Theorems \ref{Thm_opt_reg_shocks} and \ref{Thm_compactness_shocks} see \cite{ReintjesTemple_RT_shocks}, the main ideas of proof are presented here in the case $\Gamma, d\Gamma \in W^{m,p}$, $m\geq 1$, $p>n$.

\section{Applications to General Relativity}  \label{Sec_Applications}

The Einstein equations $G=\kappa T$ of General Relativity are covariant tensorial equations defined independent of coordinates.  The unknowns in the equations are the gravitational metric tensor $g$ coupled to the variables which determine the sources in $T$.  The existence of solutions, {\it apriori} estimates, and regularity results for the Einstein equations are established by PDE methods upon choosing a suitable coordinate system (or gauge condition) in which the Einstein equations take on a solvable form \cite{Choquet}.

\subsection{Application of Uhlenbeck compactness to the Einstein equations} \label{Sec_appl_Einstein-eqns}

The difficult part in a convergence proof for a PDE existence theory is typically the problem of establishing a uniform bound on the highest order derivatives, sufficient to apply Sobolev compactness. For connections, Uhlenbeck compactness tells us that it is enough to establish a bound on just the Riemann curvature, not all highest order derivatives. As an application of Uhlenbeck compactness, we have the following corollary of Theorem \ref{Thm_compactness_shocks} which provides a new compactness theorem applicable to vacuum solutions of the Einstein equations when $\Gamma, \Riem(\Gamma) \in L^\infty$.

\begin{Corollary} \label{Cor_compactness}  
Let $g_i$ be a sequence of Lipschitz continuous metrics given in some coordinate system $x$, and let $\Gamma_i$ denote the Christoffel symbol of $g_i$ for each $i \in \mathbb{N}$. Assume that each $g_i$ is an approximate solution of the vacuum Einstein equations in the sense that in $x$-coordinates  $\lim\limits_{i\rightarrow \infty} \|{\rm Ric}(g_i)\|_{L^\infty} = 0.$  Assume further that there exists some constant $M>0$ such that 
\beq \label{Cor_compct_eqn1}
\|g_i\|_{L^\infty} + \|\Gamma_i\|_{L^\infty} + \|{\rm Weyl}(g_i)\|_{L^\infty} \leq M,
\eeq
where the norms are taken in $x$-coordinates and ${\rm Weyl}(g_i)$ denotes the Weyl curvature tensor of $g_i$. Assume $|\det(g_i)|$ is uniformly bounded away from zero and let $p>n$. Then there exists a subsequence of $(g_i)_{i\in \mathbb{N}}$ which converges in $x$-coordinates component-wise and weakly in $W^{1,p}$ to some metric $g$ solving the vacuum Einstein equations, ${\rm Ric}(g)=0$, and satisfying the bound \eqref{Cor_compct_eqn1}. Furthermore, by Theorem \ref{Thm_opt_reg_shocks}, there exists a coordinate transformation $x\to y$ in $W^{2,2p}$ which lifts the components of $g$ to $W^{2,p}$ and these are the $W^{2,p}$ limits of $g_{y_i}$, the components of $g_i$ in optimal coordinates $y_i$ as in (ii) of Theorem \ref{Thm_compactness}.
\end{Corollary} 

\Proof
Since the Ricci tensor together with the Weyl tensor comprise the Riemann curvature tensor, Theorem \ref{Thm_compactness_shocks} applies and yields existence of a convergent subsequence of $g_{y_i}$, the components of $g_i$ in coordinates $y_i$ of optimal regularity expressed as functions over $x$-coordinates (as in (ii) of Theorem \ref{Thm_compactness}). The convergence $g_{y_i}\to g_y$ is weakly in $W^{2,p}$ and strongly $W^{1,p}$. The resulting strong convergence of the connections $\Gamma_{y_i}$ together with $Ric(\Gamma)$ being linear in derivatives of $\Gamma$, implies that you can pass the weak limit through the curvature to show that ${\rm Ric}(g_i)$ converges to ${\rm Ric}(g)$ weakly in $L^p$ and conclude that $Ric(g)=0$, since $\lim\limits_{i\rightarrow \infty} \|{\rm Ric}(g_i)\|_{L^\infty} =0$.  See \cite{ReintjesTemple_RT_shocks} for a detailed proof and a refined statement.
\QED

Note that without Theorem \ref{Thm_compactness_shocks}, the uniform $L^\infty$ bound on a sequence of metric connections and their curvatures would only imply weak $L^p$ convergence of a subsequence of the metric connections and curvatures. But one cannot in general pass weak limits through nonlinear functions like products  \cite[Chapter 16]{Dafermos}. As a result, even though the Ricci tensor would correctly converge to zero, the limit Ricci tensor would in general not be the Ricci tensor of the limit connection, and thus the limit metric would in general fail to be a solution of the vacuum Einstein equations. 
Corollary \ref{Cor_compactness} leads one to anticipate that existence theorems for the Einstein equations might be easier to prove in coordinate systems in which the metric is non-optimal, because in coordinates where the metric is one order less smooth, the equations need impose fewer constraints and controlling the Weyl curvature tensor suffices for iteration schemes to converge. This is the case for the solutions we now consider in Application \ref{Sec_Appl_SSC}.

\subsection{Optimal regularity in spherically symmetric spacetimes} \label{Sec_Appl_SSC}

The existence theory based on Glimm's method in \cite{GroahTemple} establishes (weak) shock wave solutions of the Einstein Euler equations, which couple the unknown metric $g_{ij}$ to the unknown density $\rho$, pressure $p$ and velocity $u$ of a perfect fluid via $T^{ij}=(\rho+p)u^iu^j+pg^{ij}$ in $G=\kappa T$. The spacetime metrics of these solutions are non-optimal with curvature in $L^\infty$, but optimal metric regularity would be required to introduce locally inertial frames and geodesic curves by standard method. In this section we introduce a corollary of Theorems \ref{ThmMainCor} and \ref{Thm_opt_reg_shocks} which establishes for the first time that solutions of the Einstein equations constructed in Standard Schwarzschild Coordinates, (which have a long history in General Relativity), including the Lipschitz continuous metrics associated with shock waves in \cite{GroahTemple}, can always be smoothed to optimal regularity by coordinate transformation. The result applies at every level of regularity, but we the result here at the low regularity of GR shock waves, $\Gamma, {\rm Riem}(\Gamma) \in L^\infty$.
 
For this consider a metric in Standard Schwarzschild Coordinates (SSC)
\begin{eqnarray}
ds^2=-B(t,r)dt^2+\frac{dr^2}{A(t,r)}+r^2d\Omega^2,
\end{eqnarray}
for which the Einstein equations take on a very simple form, with the first three equations being 
\begin{eqnarray}\label{firstorder1}
-r A_r +(1-A)&=&\kappa B T^{00}r^2 \\ \label{firstorder2}
A_t &=&\kappa B T^{01}r \\ \label{firstorder3}
r\frac{B_r}{B}-\frac{1-A}{A}&=&\frac{\kappa }{A^2}T^{11}r^2.
\end{eqnarray}
From \eqref{firstorder1} - \eqref{firstorder3} we find that the metric can generically be only one level more regular than the curvature tensor, at every level of regularity, and is hence non-optimal. (See \cite{GroahTemple} for the full system of equations.) An application of Theorem \ref{Thm_opt_reg_shocks} to spherically symmetric solutions in SSC, is the following result which establishes that shock wave solutions of the Einstein equations constructed by the Glimm scheme are one order more regular than previously known \cite{GroahTemple}.

\begin{Corollary} \label{Cor_opt}
Let $T\in L^\infty$ and assume $g\equiv(A,B)$ is a (weak) solution of the Einstein equations in SSC satisfying $g\in C^{0,1}$ and hence $\Gamma\in L^\infty$ in an open set $\Omega$. Then for any $p>4$ and any $q\in\Omega$ there exists a coordinate transformation $x\to y$ defined in a neighborhood of $q$, such that, in $y$-coordinates, $g\in W^{2,p},$ $\Gamma\in W^{1,p}$.
\end{Corollary}

\Proof
In SSC the Ricci and Riemann curvature tensor have the same regularity (as can be verified using Mathematica). So $T$ in $L^\infty$ implies $d\Gamma$ in $L^\infty$, and Theorem \ref{Thm_opt_reg_shocks}  implies the corollary. 
\QED

\section{Euclidean Cartan calculus for matrix valued differential forms} \label{Sec_Prelim}

Our motivation in \cite{ReintjesTemple_ell1} for introducing matrix valued differential forms begins by expressing the Riemann curvature tensor as matrix valued $2$-form,
\beq \label{Riemann_2-form} 
 {\rm Riem}(\Gamma) = d \Gamma + \Gamma \wedge \Gamma,
\eeq
interpreting the connection $\Gamma$ as the matrix valued $1$-form $\Gamma^\mu_\nu \equiv \Gamma^\mu_{\nu i} dx^i$. By a matrix valued differential $k$-form $A$ we mean an $(n\times n)$-matrix whose components are $k$-forms over the spacetime region $\Omega\subset \R^n$, and we write
\beq \label{def_matrix_valued_diff}
A \equiv \sum_{i_1< ... < i_k} A_{i_1...i_k} dx^{i_1} \wedge ... \wedge dx^{i_k},
\eeq 
for $(n\times n)$-matrices  $A_{i_1...i_k}$, assuming total anti-symmetry in the indices $i_1,...,i_k \in \{1,...,n\}$.  The wedge product of $A$ with a matrix valued $l$-form $B = B_{j_1...j_l} dx^{j_1} \wedge ... \wedge dx^{j_l}$ is then defined by  
\begin{eqnarray} \label{def_wedge}
A \wedge B  
&\equiv & \frac{1}{l!k!} A_{i_1...i_k} \mm B_{j_1...j_l} \; dx^{i_1} \wedge ... \wedge dx^{i_k} \wedge dx^{j_1} \wedge ... \wedge dx^{j_l}, 
\end{eqnarray}
where ``$\cdot$'' denotes standard matrix multiplication. So $\Gamma \wedge \Gamma$ in \eqref{Riemann_2-form} is non-zero, unless all component matrices mutually commute (which would suffice for $\Gamma$ being a $1$-form). We introduce the matrix valued inner product
\beq \label{def_inner-product}
\langle A\; ; B \rangle^\mu_\nu \equiv \sum_{i_1<...<i_k} A^\mu_{\sigma\: i_1...i_k} B^\sigma_{\nu\: i_1...i_k} ,
\eeq 
which is a matrix valued version of the Euclidean inner product of $k$-forms, and the Hodge star operator $*$ by
\beq \label{def_Hodge}
A \wedge (*B)   \equiv \langle A\: ;B\rangle  dx^1 \wedge ... \wedge dx^n .
\eeq
The exterior derivative is defined as 
\begin{eqnarray} \label{def_exterior_deriv}
d A &\equiv &  
\partial_l A_{[i_1...i_k]} dx^l \wedge dx^{i_1} \wedge ... \wedge dx^{i_k} ,
\end{eqnarray}
the co-derivative as the $(k-1)$-form 
\beq \label{def_coderiv}
\delta A \equiv  (-1)^{(k+1)(n-k)} * \big( d (* A) \big)
\eeq
and the Laplace operator as 
\beq \label{def_Laplacian}
\Delta  \equiv \delta d + d \delta.
\eeq
The derivative operations \eqref{def_exterior_deriv}, \eqref{def_coderiv} and \eqref{def_Laplacian} act on matrix components separately and behave like the analogous operations on scalar valued differential forms. In particular, c.f. Theorem 3.7 in \cite{Dac}, $\Delta$ acts component-wise as the Euclidean Laplacian, 
\beq \label{basics_Laplacian}
(\Delta A)^\mu_{\nu i_1...i_k} 
= \Delta \big(A^\mu_{\nu i_1...i_k}\big) 
= \sum_{j=1}^n \partial_{j}\partial_{j}\big(A^\mu_{\nu i_1...i_k}\big).
\eeq 

We convert matrix valued differential forms to vector valued forms as follows: We let an arrow over a matrix valued $0$-form $A$ denote the conversion of $A$ into its equivalent vector valued $1$-form, i.e., 
\beq \label{defnarrow1}
\vec{A}\equiv A^\alpha_i dx^i,
\eeq 
where $\alpha$ labels the components of the vector. By this, we express the integrability condition for the Jacobian $J$ as $d\vec{J} =0$, since                
\beq \nonumber 
Curl(J)\equiv \frac12 \big( J^\alpha_{i,j} - J^\alpha_{j,i} \big) dx^j \otimes dx^i = J^\alpha_{i,j} dx^j\wedge dx^i =  d(J^\alpha_i dx^i) \equiv  d\vec{J}^\alpha.
\eeq 
Moreover, for a matrix valued $k$-form $A$, we define the operation 
\beq \label{Def_vec-div}
\overrightarrow{\text{div}}(A)^\alpha \equiv \sum_{l=1}^n \partial_l \big( (A^\alpha_l)_{i_1,,,i_k}\big) dx^{i_1}\wedge . . . \wedge dx^{i_k} ,
\eeq
which creates a vector valued $k$-form.  The operations \eqref{defnarrow1} - \eqref{Def_vec-div} are meaningful when the dimension of the matrices equals the dimension of the physical space. For the proof of Theorems \ref{Thm_main} and \ref{ThmMain} we extend  in \cite{ReintjesTemple_ell1} various identities of classical Cartan calculus to the setting of matrix valued differential forms. The key identity required to close the RT-equations within the appropriate regularity classes is the identity
\beq \label{regularity-miracle}
d \big(\overrightarrow{\delta ( J \mm \Gamma )}\big) 
= \overrightarrow{\text{div}} \big(dJ \wedge \Gamma\big) + \overrightarrow{\text{div}} \big( J\mm d\Gamma\big) ,
\eeq
which applies to matrix valued $1$-forms $\Gamma$ and matrix valued $0$-forms $J$, c.f. \cite{ReintjesTemple_ell1} for proofs. This identity has no analogue for classical scalar valued differential forms.

\section{The Riemann-flat condition} \label{Sec_Riemann-flat}

To begin, consider the transformation law for a connection  
\beq \label{prelim_connection_transfo_1}
(J^{-1})^k_\alpha \big( \partial_j J^\alpha_{i} + J^\beta_i J^\gamma_j \Gamma^\alpha_{\beta\gamma} \big) = \Gamma^k_{ij} ,
\eeq
where $\Gamma^k_{ij}$ denotes the components of the connection in $x^i$-coordinates,  $\Gamma^\alpha_{\gamma\beta}$ denotes its components in $y^{\alpha}$-coordinates and $J^\alpha_i \equiv \frac{\partial y^\alpha}{\partial x^i}$. 
Assume now that $\Gamma^k_{ij} \in W^{m,p}(\Omega)$, $\Gamma^\alpha_{\gamma\beta} \in W^{m+1,p}(\Omega)$ and $J^\alpha_i \in W^{m+1,p}(\Omega)$, for $m\geq 1$. In other words, assume the Jacobian $J$ smooths the connection $\Gamma^k_{ij}$ by one order.  For these given coordinates $x$ and $y$, we introduce
\beq \label{def_tensor_Gammati}
\tilde{\Gamma}^k_{ij} \equiv (J^{-1})^k_\alpha  J^\beta_i J^\gamma_j \Gamma^\alpha_{\beta\gamma},
\eeq
which defines a field in $x$-coordinates. By imposing that $\Gammati^k_{ij}$ should transform as a $(1,2)$-tensor, \eqref{def_tensor_Gammati} defines a {\it tensor} $\Gammati$.  Now, \eqref{prelim_connection_transfo_1} can be written equivalently as 
\beq \label{prelim_connection_transfo_2}
(J^{-1})^k_\alpha \; \partial_j J^\alpha_{i} = (\Gamma-\tilde{\Gamma})^k_{ij},
\eeq
which we interpret as a condition on the fields $J$ and $\Gammati$ in $x$-coordinates. To obtain the Riemann-flat condition from \eqref{prelim_connection_transfo_2}, observe that adding a tensor to a connection yields another connection, so \eqref{prelim_connection_transfo_2} is the condition that $J$ transforms the connection $\Gamma-\tilde{\Gamma}$ to zero.  This implies that $\Gamma-\tilde{\Gamma}$ is a Riemann-flat connection, ${\rm Riem} (\Gamma-\tilde{\Gamma}) =0$.  We conclude, that the existence of a coordinate transformation $x\to y$ which lifts the connection regularity by one order implies the {\it Riemann-flat condition}, that is, the condition that there exists a symmetric $(1,2)$-tensor $\Gammati$ one order more regular than $\Gamma$ such that ${\rm Riem} (\Gamma-\tilde{\Gamma}) =0$. The following theorem records several further equivalences which, in particular, imply that the inverse implication is also true. 

\begin{Thm} \label{Thm_Prelim}
Let $\Gamma^k_{ij}$ be a symmetric\footnote{We remark that our main results in Section \ref{Sec_Results} do not require assuming symmetry of $\Gamma$ or $\Gammati$.} connection in $W^{m,p}(\Omega)$ for $m\geq 1$ and $p>n$  (in coordinates $x^i$). Then the following statements are equivalent: 
\vspace{-.4cm}
\begin{enumerate}[(i)]
\item There exists a coordinate transformation $x^i\to y^{\alpha}$ with Jacobian $J \in W^{m+1,p}(\Omega)$ such that $\Gamma^\alpha_{\beta\gamma}\in W^{m+1,p}(\Omega)$ in $y$-coordinates.
\item There exists a symmetric $(1,2)$-tensor $\Gammati \in W^{m+1,p}(\Omega)$ and a matrix field $J \in W^{m+1,p}(\Omega)$ solving
\begin{align} 
J^{-1}dJ =\Gamma-\tilde{\Gamma},& \label{Riemann_flat_J-form_prelim} \\
{\rm Curl}(J)^\alpha_{ij} \equiv  J^\alpha_{i,j} - J^\alpha_{j,i} = 0.& \label{J_integrability_prelim} 
\end{align}
\item There exists a symmetric $(1,2)$ tensor $\Gammati \in W^{m+1,p}(\Omega)$ such that $\Gamma-\Gammati$ is Riemann-flat: ${\rm Riem} (\Gamma-\tilde{\Gamma}) =0$.
\item There exists a symmetric $(1,2)$-tensor $\Gammati \in W^{m+1,p}(\Omega)$ which, when viewed as a matrix valued $1$-form in $x$-coordinates,  solves
\beq \label{Riemann_flat_curl-form_prelim}
d\tilde{\Gamma} = d\Gamma + \big(\Gamma - \tilde{\Gamma}\big) \wedge \big(\Gamma - \tilde{\Gamma}\big) .
\eeq
\end{enumerate}
\end{Thm}

\Proof
Note that \eqref{Riemann_flat_J-form_prelim} is a restatement of \eqref{prelim_connection_transfo_2} in the formalism of matrix valued differential forms, and \eqref{J_integrability_prelim} is the condition that $J$ is integrable to define a coordinate system, c.f. \cite{ReintjesTemple_geo}. This shows that (i) and (ii) are equivalent and that (ii) implies (iii). The equivalence of (iii) and (iv) follows from the expression  of the Riemann tensor as a matrix valued $2$-form in \eqref{Riemann_2-form}. Finally, the implication (iii) to (i) is proved in \cite{ReintjesTemple_geo} when $\Gamma\in L^{\infty}$ and $\tilde{\Gamma},J\in C^{0,1}$. The more regular case here, $\Gamma\in W^{m,p}$ and $\tilde{\Gamma},J\in W^{m+1,p}$, follows by the analogous argument without mollification.   
\QED

By Theorems \ref{Thm_Prelim} and \ref{Thm_main}, existence for the RT-equations is equivalent to the Riemann-flat condition. Thus, as an immediate application of Theorems \ref{Thm_main} and \ref{ThmMain}, we obtain the following analog of a Nash-type embedding theorem for connections with discontinuities in the $m$'th derivatives:   

\begin{Corollary} \label{Cor_Nash}   
If $\Gamma,Riem(\Gamma)\in W^{m,\infty}(\Omega)$, {\rm(}so $\Gamma\in W^{m,p}_{loc}$ for $p>n$, and $\Gamma$ can be taken to have bounded discontinuities in the $m$'th derivatives$)$, then in a neighborhood of each point in $\Omega$ there exists a Riemann-flat connection $\hat{\Gamma}$, $($namely, $\hat{\Gamma}=\Gamma-\tilde{\Gamma}'${\rm )}, which contains discontinuities in the $m$'th derivatives at the same locations as $\Gamma$, and these discontinuities are the same to within the addition of a continuous function.   
\end{Corollary} 

For example, for BV $m$'th derivatives, the flat connection $\hat{\Gamma}$ would have the same jumps at all jump discontinuities of $\Gamma$, \cite{Smoller,Dafermos}. Theorem \ref{Thm_Prelim} implies that to prove optimal regularity it would suffice to construct a Nash-type extension of the singular set of $\Gamma$ to a Riemann-flat connection. Establishing optimal regularity by the RT-equations turns out to be more feasible.

\section{Proof of Theorem \ref{Thm_main}}   \label{Sec_Proof_Thm1}

We begin by outlining the ideas and steps in the derivation of the RT-equations set out in detail in \cite{ReintjesTemple_ell1}. The idea is that by Theorem \ref{Thm_Prelim} the Riemann-flat condition ${\it Riem}(\Gamma-\Gammati)=0$ gives the equation \eqref{Riemann_flat_curl-form_prelim}, namely, $d\Gammati=d\Gamma+(\Gamma-\Gammati)\wedge(\Gamma-\Gammati)$, which we view as an equation for $\Gammati$. This can be augmented to a first order system of Cauchy-Riemann equations by addition of an equation for $\delta\Gammati$ with arbitrary right hand side.  But to obtain a solvable system, we couple this Cauchy-Riemann system in the unknown $\Gammati$, to equation \eqref{Riemann_flat_J-form_prelim}, namely $J^{-1}dJ=\Gamma-\Gammati$, for the unknown Jacobian $J$. But equations \eqref{Riemann_flat_curl-form_prelim} and \eqref{Riemann_flat_J-form_prelim} are not independent, since both are equivalent to the Riemann-flat condition. To obtain two independent equations, we employ the identity $d\delta+\delta d=\Delta$ to derive two semi-linear elliptic Poisson equations, one for $\Delta\Gammati$ and one for $\Delta J$. This results in the two second order equations \eqref{eqn1} - \eqref{eqn2}, which closes in $(J,\Gammati)$ for fixed $A$ upon setting $\delta \Gammati = J^{-1}A$. The equations are formally correct at the levels of regularity sufficient for $J$ and $\Gammati$ to be one order smoother than $\Gamma$, consistent with known results on elliptic smoothing by the Poisson equation in $L^p$-spaces, \cite{Dac,Evans,GilbargTrudinger}. 

To impose the integrability condition for $J$, we use the freedom in $\delta\Gammati$ to interpret $A$ as a {\it variable} on the right hand side of \eqref{eqn1}  and \eqref{eqn2}, and impose $Curl(J)=0$ by asking that $A$ solve the equation obtained by requiring $d$ of the vectorized right hand side of the $J$ equation \eqref{eqn2} to equal zero. When taking $d$ of the right hand side of \eqref{eqn2}, we encounter the term $d \big(\overrightarrow{\delta ( J \mm \Gamma )}\big)$ which seems to involve uncontrolled derivatives on $\Gamma$, hence one derivative too low to get the required regularity $A\in W^{m,p}$.\footnote{Note, $A\in W^{m,p}$ is needed for \eqref{eqn1} - \eqref{eqn2} to imply the required regularity for $(J,\Gammati)$.} But, surprisingly, this term can be re-expressed in terms of $d\Gamma$ by the fortuitous identity \eqref{regularity-miracle}, so this term is in fact one order smoother than it initially appears to be. (This confirms that our assumptions need only control $d\Gamma$ in $W^{m,p}$, but not the complementary derivatives $\delta\Gamma$, which, by (\eqref{Gaffney}), measure the derivatives not controlled by $d\Gamma$.) This gives \eqref{eqn3}. The final form of the RT-equations is then obtained by augmenting \eqref{eqn1} - \eqref{eqn3} by equation \eqref{eqn4}. This represents the ``gauge freedom'' to impose $\delta A=v$. This completes the derivation of the RT-equations and establishes the backward implication in Theorem \ref{Thm_main}.\footnote{One might wonder why we were not able to obtain an equation for the coordinate transformation $y$ directly, so that the simpler $dy=J$ would replace the integrability condition $Curl(J)=0$.   This is because, starting with the Riemann-flat condition $Riem(\Gamma-\Gammati)=0$, the gauge freedom enters through the freedom to impose $\delta\tilde{\Gamma}$, and this expresses itself in the additional variable $A$ on the right hand side of equation (\ref{eqn1}). To close the system, we then need a differential equation for $A$, which naturally comes from $Curl(J)=d\vec{J}=0$ by setting $d$ of the vectorized right hand side of (\ref{eqn2}) equal to zero, leading to the equation (\ref{eqn3}) for $A$.  Thus to obtain a closed solvable system, we are essentially forced to impose the integrability condition on $J$ in the form $Curl\vec{J}=0$.}

We now outline the proof of the forward implication in Theorem \ref{Thm_main}, namely, that a solution of the RT-equations produces a Jacobian $J$ which lifts $\Gamma$ to optimal regularity. So assume $(J,\Gammati,A)$ solves the RT-equations. We first show that $J$ is integrable to coordinates. If $J$ is a solution of \eqref{eqn2} and $A$ solves \eqref{eqn3} - \eqref{eqn4} with boundary data \eqref{bdd1}, then, as shown in \cite{ReintjesTemple_ell1}, $\Delta(d\vec{J})=0$ in $\Omega$. Thus, since $d\vec{J}$ is assumed to vanish on $\partial\Omega$ by \eqref{bdd1}, it follows that the harmonic form $d\vec{J}$ is zero everywhere in $\Omega$, so $J$ is integrable to coordinates. To complete the forward implication, note that $\Gammati$ need not satisfy the Riemann-flat condition because the RT-equations have a larger solution space than the first order equations from which they are derived. So we define $\Gammati' \equiv \Gamma-J^{-1}dJ$, which meets the Riemann-flat condition by \eqref{Riemann_flat_J-form_prelim}. But an additional argument is required to show that $\Gammati'$, like $\Gammati$, is indeed one level smoother than $\Gamma$, as stated in Lemma \ref{Lemma_main}. For this, we use \eqref{eqn1} - \eqref{eqn2} to show  $\Delta \Gammati' \in W^{m-1,p}(\Omega)$, (by deriving equation (4.32) in \cite{ReintjesTemple_ell1}), so that standard estimates of elliptic regularity theory imply the desired smoothness $\Gammati' \in W^{m+1,p}(\Omega')$ on any compactly contained subset $\Omega'$ of $\Omega$, (c.f. \cite{ReintjesTemple_ell1}). This establishes the forward implication in Theorem \ref{Thm_main}.

In summary, we start with two equivalent first order equations, one for $d\Gammati$ and one for $dJ$,  both equivalent to the Riemann-flat condition.   Out of these we create two independent nonlinear Poisson equations in $\Gammati$ and $J$ which have a larger solution space. The resulting system has the freedom to impose an auxiliary solution $A$ through the gauge freedom to impose $\delta\Gammati$.  Since the solution space is larger, not all solutions of the RT-equations provide a $\Gammati$ which solves the Riemann-flat condition, but given any solution $(\Gammati,J,A)$ of \eqref{eqn1} - \eqref{bdd1}, we show that there is enough freedom in $A$ so that there always exists $A'$ such that $(\Gammati',J,A'),$ solves the RT-equations with $\Gammati'=\Gamma-J^{-1}dJ$. Then $\Gammati'$ meets the Riemann-flat condition by construction, and $J$ is the Jacobian of a coordinate transformation which takes $\Gamma$ to optimal regularity.

\section{Proof of Theorem \ref{ThmMain}}    \label{Sec_Proof_Thm2}

The biggest challenge of this research program was to discover a system of nonlinear equations for optimal regularity, the RT-equations, and formulate them so that existence of solutions to the nonlinear equations could be deduced from known theorems of elliptic regularity theory.  The existence proof in \cite{ReintjesTemple_ell2}, which we outline in this section, was the first to demonstrate that obtaining optimal regularity by the RT-equations works. The extension of this existence theory to include $L^\infty$ connections in \cite{ReintjesTemple_RT_shocks} then demonstrates that the method of obtaining optimal regularity via the RT-equations is efficient enough to remove apparent singularities in such connections.

The strategy of proof here is to deduce convergence of an iteration scheme for approximating the nonlinear equations, from two standard theorems on the Dirichlet  problem (stated below) of the linear theory of elliptic PDE's in $L^p$ spaces, \cite{Dac,GilbargTrudinger,Evans}. To begin, we rewrite the RT-equations \eqref{eqn1} - \eqref{eqn4} in the following compact form
\begin{eqnarray} 
\Delta \Gammati &=& \tilde{F}(\Gammati,J,A), \label{eqn11} \\
\Delta J &=& F(\Gammati,J) - A , \label{eqn22} \\
d \vec{A} &=& d\vec{F}(\Gammati,J)  \label{eqn33}\\
\delta \vec{A} &=& v,  \label{eqn44}
\end{eqnarray}
where $\vec{F}(\Gammati,J)$ is the vectorization of $F(\Gammati,J)$ and where 
\begin{eqnarray} \nonumber
\tilde{F}(\Gammati,J,A) & \equiv &  \delta d \Gamma - \delta \big( d(J^{-1})\wedge dJ\big) + d(J^{-1} A ), \cr
F(\Gammati,J) & \equiv &  \delta ( J \mm \Gamma ) - \langle d J ; \tilde{\Gamma}\rangle,
\end{eqnarray}
(so $d\vec{F}(\Gammati,J)$ equals the right hand side of \eqref{eqn3}, c.f. the derivation of (3.40) in \cite{ReintjesTemple_ell1}). Note (\ref{eqn33}) - (\ref{eqn44}) take the Cauchy-Riemann form $d\vec{A}=f$, $\delta\vec{A}=g$. The consistency conditions $df=0$, $\delta g=0$ are met, since the right hand side of \eqref{eqn33} is exact and since $\delta v=0$ holds as an identity for $0$-forms. 

To handle the nonlinearities in \eqref{eqn11} - \eqref{eqn44}, we introduce a small parameter $\epsilon>0$ below by using the freedom to restrict to small neighborhoods, and then apply linear elliptic estimates in $L^p$ spaces to establish convergence at the sought after levels of regularity for sufficiently small $\epsilon>0$. But we still have the problem of how to handle the non-standard boundary condition (\ref{bdd1}), which is neither Neumann nor Dirichlet data for (\ref{eqn22}).   We now introduce an equivalent formulation of the boundary condition (\ref{bdd1}) for (\ref{eqn22}), which has the advantage that it reduces to standard Dirichlet data for $J$ at each stage of the iteration scheme below. For this, observe that (\ref{eqn33}) implies the consistency condition $d\big(\vec{F}(\Gammati,J)-\vec{A}\big)=0$, so that one can solve 
\beq \label{definepsi}
\begin{cases}
d\Psi =\vec{F}(\Gammati,J)-\vec{A}, \\
\delta\Psi = 0,
\end{cases}
\eeq
for a vector valued $0$-form $\Psi$, (c.f. Theorem 7.4 in \cite{Dac}). Next, let $y$ be any solution of  
\begin{eqnarray}\label{definey}
\Delta y=\Psi.
\end{eqnarray}
We now claim that in place of the Poisson equation (\ref{eqn2}) for $J$ with the boundary condition (\ref{bdd1}), it suffices to solve \eqref{eqn2} with boundary data
\beq
\vec{J}=dy\ \ {\rm on}\ \ \partial\Omega. \label{thepoint2}
\eeq 
To see this, write $\Delta dy=d\Delta y=d\Psi=\vec{F}-\vec{A}=\Delta\vec{J}$,
which uses that, after taking $vec$ on both sides of the $J$-equation \eqref{eqn2}, the operation $vec$ commutes with $\Delta$ on the left hand side of \eqref{eqn2} because the Laplacian acts component-wise. 
Thus, $\Delta(\vec{J}-dy)=0$ in $\Omega$ and $\vec{J}-dy=0$ on $\partial\Omega$, which implies by uniqueness of solutions of the Laplace equation that $\vec{J}=dy$ in $\Omega$. Since second derivatives commute, we conclude that $d\vec{J}=Curl(J)=0$ in $\Omega$, on solutions of \eqref{eqn2} with boundary data (\ref{thepoint2}), as claimed.   The point of using (\ref{thepoint2}) in place of (\ref{bdd1}) is that (\ref{thepoint2}) is standard Dirichlet data for $J$ in the following iteration scheme.  

We now discuss the iteration scheme introduced in \cite{ReintjesTemple_ell2} for approximating solutions of the RT-equations \eqref{eqn11} - \eqref{eqn44}. To start, assume a given connection $\Gamma \in W^{m,p}$ defined in $x$-coordinates on a bounded and open set $\Omega \subset \R^n$ with smooth boundary. We take $v=0$ in \eqref{eqn4} to fix the freedom to choose $v \in W^{m-1,p}(\Omega)$. For the existence proof, we define a sequence of differential forms $(A_{k},\Gammati_{k},J_{k})$ in $\Omega$, and prove convergence to a solution  $(A,\Gammati ,J)$ of \eqref{eqn11} - \eqref{eqn44} with boundary data \eqref{bdd1} in the limit $k\to\infty$.  Define the iterates $(A_{k},\Gammati_{k},J_{k})$ by induction as follows: To start, take $J_0$ to be the identity matrix and set $\Gammati_0=0$. Assume then $\Gammati_k$ and $J_k$ are given for some $k\geq 0$.  Define $A_{k+1}$ as the solution of
\beq \label{iteration_eqnA}
\begin{cases}
d\vec{A}_{k+1} = d\vec{F}(\Gammati_k,J_k), \cr
\delta \vec{A}_{k+1}=0,
\end{cases}
\eeq    
for $A_{k+1} \mm N=0$ on $\partial \Omega$, where $N$ is the unit normal vector of $\partial \Omega$ which is multiplied by the matrix $A_{k+1}$.  To introduce the Dirichlet data for $J_{k+1}$, we first define the auxiliary variables $\psi_{k+1}$ and $y_{k+1}$, as the solutions of 
\beq \label{iteration_eqnPsi}
d\Psi_{k+1}=\vec{F}(\Gammati_k,J_k)-\overrightarrow{A_{k+1}}, 
\eeq
with $\int_\Omega \Psi_{k+1} dx =0$ (playing the role of boundary data for the gradient system \eqref{iteration_eqnPsi}), and  
\beq \label{iteration_eqn_y}
\Delta y_{k+1}=\Psi_{k+1}
\hspace{1cm}\text{with} \hspace{1cm}
y_{k+1}(x) = x \ \ {\rm on}\ \ \partial\Omega.
\eeq
Now define $J_{k+1}$ to be the solution of the following standard {\it Dirichlet} boundary value problem,   
\begin{eqnarray} \label{iteration_eqn_J}
\Delta J_{k+1} = F(\Gammati_{k},J_k)-\overrightarrow{A_{k+1}},
\hspace{1cm}\text{with} \hspace{1cm}
\overrightarrow{J_{k+1}} = dy_{k+1} \ \ {\rm on}\ \ \partial\Omega; 
\end{eqnarray} 
and define $\Gammati_{k+1}$ as the solution of \footnote{Note, since we only need to construct a particular solution, essentially any boundary condition could be chosen for \eqref{iteration_eqnPsi}, \eqref{iteration_eqn_y} and \eqref{iteration_eqn_Gammati}. Note also, that the definitions of $A_{k+1}$, $\Psi_{k+1}$ and $y_{k+1}$ do not require the previous iterates $A_k$, $\Psi_{k}$ and $y_{k}$.}
\beq \label{iteration_eqn_Gammati}
\Delta \Gammati_{k+1} = \tilde{F}(\Gammati_k,J_k,A_{k+1}),
\hspace{1cm}\text{with} \hspace{1cm}
\Gammati_{k+1} =0  \ \ {\rm on}\ \ \partial\Omega.
\eeq

To prove that there exists a well-defined sequence of iterates $(J_k,\Gammati_k,A_k)$, $k\in \mathbb{N}$, and establish convergence, we introduce a small parameter $\epsilon>0$. For this, let $\Gamma^*$ be a connection in $x$-coordinates satisfying        
\begin{eqnarray}
\|(\Gamma^*, d\Gamma^*)\|_{W^{m,p}(\Omega)} \equiv \|\Gamma^*\|_{W^{m,p}(\Omega)} \: + \: \|d\Gamma^*\|_{W^{m,p}(\Omega)} = C_0 \leq M, \label{Gamma-bound}
\end{eqnarray} 
for $m\geq 1$ and some constant $C_0 >0$ independent of $\epsilon$. To introduce the small parameter assume that $\Gamma$ scales with $\epsilon>0$ according to
\begin{eqnarray}
\Gamma=\epsilon\; \Gamma^*. \label{small_Gamma}
\end{eqnarray} 
Note that assumptions \eqref{Gamma-bound} and \eqref{small_Gamma} can be made without loss of generality regarding the \emph{local} problem of optimal metric regularity. To see this assume that $\Omega$ is the ball of radius $1$. Then given any connection $\Gamma'(y) \in W^{m,p}(\Omega)$ with $\|(\Gamma', d\Gamma')\|_{W^{m,p}(\Omega)} \leq M$, we can define $\Gamma^*(x)$ as the restriction of $\Gamma'$ to the ball of radius $\epsilon$ with its components transformed as scalars to the ball of radius $1$ by the transformation $y=\epsilon x$. We then define $\Gamma(x)$ as the connection resulting from transforming $\Gamma'(y)$ \emph{as a connection} under the coordinate transformation $y=\epsilon x$. This establishes the scaling \eqref{small_Gamma} together with the bound  $\|(\Gamma^*, d\Gamma^*)\|_{W^{m,p}(\Omega)} \leq M$, which for ease of presentation we take here as the equality in \eqref{Gamma-bound} in terms of $C_0>0$. We conclude that, given any connection $\Gamma'$, local existence of a solution of the RT-equations with $\Gamma=\Gamma'$ follows from the existence of a solution of the RT-equations with $\Gamma=\epsilon\Gamma^*$ and \eqref{Gamma-bound} - \eqref{small_Gamma} for some $\epsilon>0$.  Thus, without loss of generality, we assume (\ref{small_Gamma}), c.f. \cite{ReintjesTemple_ell2} for details. 

To incorporate $\epsilon$ into the RT-equations, we assume the scaling ansatz 
\beq \label{ansatz_scaling}
J_k=I+\epsilon \, J^*_k , \hspace{.3cm} 
\tilde{\Gamma}_k=\epsilon\: \tilde{\Gamma}^*_k, \hspace{.3cm} 
A_k = \epsilon A^*_k, \hspace{.3cm} u_k\equiv (J^*_k,\Gammati^*_k), \hspace{.3cm} a_k\equiv A^*_k.
\eeq 
Substitute \eqref{small_Gamma} and \eqref{ansatz_scaling} into the RT-equations \eqref{eqn11} - \eqref{eqn44} for $v \equiv 0$ and dividing by $\epsilon>0$, we obtain the following equivalent set of equations:
\beq  \label{pde}
\Delta u = F_u(u,a), \hspace{1cm} \text{and} \hspace{1cm}
\begin{cases} 
d\vec{a} = F_a(u) , \\
\delta\vec{a} = 0  ,
\end{cases}
\eeq
where \small
\begin{align}  
& F_u(u,a) \equiv  \left(\msp \begin{array}{c} \delta d \Gamma^*  -  \delta d\big( J^{-1} \mm dJ^* \big)  +  d (J^{-1} a)  \cr \delta \Gamma^*   + \epsilon\: \delta ( J^* \mm \Gamma^* ) - \epsilon\: \langle d J^* ; \tilde{\Gamma}^*\rangle - a \end{array} \msp\right) ,     \label{Def_Fu}  \\  
& F_a(u)  \equiv   \overrightarrow{\text{div}} \big(d\Gamma^*\big) + \epsilon\: \overrightarrow{\text{div}} \big( J^* \mm d\Gamma^*\big) + \epsilon\: \overrightarrow{\text{div}} \big(dJ^* \wedge \Gamma^*\big)  -   \epsilon\: d\big(\overrightarrow{\langle d J^* ; \tilde{\Gamma}^*\rangle }\big).  \label{Def_Fa} 
\end{align}        
\normalsize

\noindent Under assumption (\ref{small_Gamma}), the iterates defined by \eqref{iteration_eqnA} - \eqref{iteration_eqn_Gammati} generate corresponding iterates $(u_k,a_k)$ which successively solve \eqref{pde}, as well as iterates $\Psi_k^* = \frac{1}{\epsilon} \Psi_k$ and $y_k^* = \frac{1}{\epsilon} y_k$.  It remains to prove that $(u_k, a_k)$ and $( \Psi_k^*,y_k^*)$ are well defined and converge for $\epsilon$ sufficiently small.   We state the results in two theorems:

\begin{Thm} 
Assume $(u_k,a_k) \in W^{m+1,p}(\Omega)\times W^{m,p}(\Omega)$. Then $(u_{k+1},a_{k+1})$ is well-defined and bounded in the same Sobolev space for $\epsilon>0$ sufficiently small.
\end{Thm}

\noindent{\it Proof.}  This is implied by the following two well known theorems from linear elliptic PDE theory,\footnote{See Theorem 7.4 in \cite{Dac} and Theorem 9.15 in \cite{GilbargTrudinger} respectively, as well as the more detailed description in \cite[Ch. 2]{ReintjesTemple_ell2}.} which both extend component-wise to matrix and vector valued differential forms. (The possibility that we might reduce the existence theory to these two theorems was the guiding principle in the formulation of the RT-equations.)
\vspace{.2cm}

\noindent{\bf Theorem: (Cauchy-Riemann)}
{\it Let $f\in W^{m,p}(\Omega)$ be a $2$-form with $df=0$ and assume further that $f=d v$ for some $1$-form $v \in W^{m,p}(\Omega)$, where $m\geq 0$, $n\geq2$. Then there exists a $1$-form $u=u_i\, dx^i \in W^{m+1,p}(\Omega)$ which solves $du=f$ and $\delta u =0$ in $\Omega$ with boundary data $u \cdot N=0$ on $\partial\Omega$.  Moreover, there exists a constant $C_e>0$ depending only on $\Omega$, $m,n,p$, such that}
\beq \label{Gaffney}
\| u \|_{W^{m+1,p}(\Omega)}  \leq C_e   \| f \|_{W^{m,p}(\Omega)} .
\eeq
\vspace{.2cm}

\noindent{\bf Theorem: (Poisson)} 
{\it Let $f\in W^{m-1,p}(\Omega)$ and $u_0 \in W^{m+1,p}(\partial\Omega)$ both be scalar functions, and $m\geq 1$. Then there exists $u \in W^{m+1,p}(\Omega)$ which solves the Poisson equation $\Delta u = f$ with Dirichlet data $u|_{\partial\Omega} = u_0|_{\partial\Omega}$.  Moreover, there exists a constant $C_e>0$ depending only on $\Omega$, $m,n,p$ such that }
\beq \label{Poissonelliptic_estimate_Lp}
\| u \|_{W^{m+1,p}(\Omega)} \leq C_e \Big( \| f \|_{W^{m-1,p}(\Omega)} +  \| u_0 \|_{W^{m+1,p}(\Omega)} \Big).
\eeq
\vspace{.2cm}

\noindent Namely, putting $(u_k,a_k)$ into the right hand side of \eqref{pde}, using Morrey's inequality to estimate quadratic terms by the supnorm times appropriate Sobolev bounds, we obtain bounds of $F_a(u)$ and $ F_u(u,a)$ in suitable Sobolev norms. (Morrey's inequality states that, when $p>n$, functions $f \in W^{m,p}(\Omega)$ satisfy 
\beq \label{Morrey_textbook}
\| f \|_{C^{0,\alpha}(\overline{\Omega})}  \leq C_M \|f\|_{W^{1,p}(\Omega)},
\eeq
where $\alpha \equiv 1 - \frac{n}{p}$ and $C_M>0$ is a constant depending only on $n$, $p$ and $\Omega$, c.f. \cite[Chapter 5]{Evans}.) Combining these bounds with the above elliptic estimates \eqref{Gaffney} and \eqref{Poissonelliptic_estimate_Lp}, generates estimates for $a_{k+1} \in W^{m,p}(\Omega)$, $\Psi_{k+1}^* \in W^{m,p}(\Omega)$, $y_{k+1}^* \in W^{m+2,p}(\Omega)$ and then $u_{k+1} \in W^{m+1,p}(\Omega)$, in terms of $(u_k,a_k)$, for $\epsilon>0$ sufficiently small.  For details see \cite{ReintjesTemple_ell2}.  \hfill $\Box$        

\begin{Thm}   
There exists $(u,a)$ such that the sequence $(u_k,a_k)_{k\in \mathbb{N}}$ converges to $(u,a)$ in $W^{m+1,p}(\Omega)\times W^{m,p}(\Omega)$  as $k\to\infty$, and $(u,a)$ solves \eqref{pde}. 
\end{Thm}

\noindent{\it Proof.} 
In order to establish convergence of the sequence of iterates $(u_k,a_k)_{k\in \mathbb{N}}$ in $W^{m+1,p}(\Omega)\times W^{m,p}(\Omega)$, we require estimates on the differences $\overline{a_{k}} \equiv  a_k- a_{k-1}$ and $\overline{u_{k}} \equiv  u_{k}-u_{k-1}$, in terms of the corresponding differences of source terms, $\overline{F_{u}(u_{k},a_{k+1})} \equiv  F_{u}(u_{k},a_{k+1}) - F_{u}(u_{k-1},a_{k})$ and $\overline{F_a(u_{k})} \equiv  F_a(u_{k}) - F_a(u_{k-1})$.  Combining the elliptic estimate \eqref{Gaffney} and \eqref{Poissonelliptic_estimate_Lp} with source estimates, (for which we use that $W^{m,p}$ is closed under multiplication when $m\geq 1$ and $p>n$, by Morrey's inequality), the main estimate proven in \cite{ReintjesTemple_ell2} is the following Sobolev space estimate which holds for $\epsilon>0$ sufficiently small: 

\begin{Lemma} \label{Lemma_nonlinear_estimate}
Assume $\epsilon\leq  \min \big(\epsilon(k),\epsilon(k-1)\big)$, where $\epsilon(k) \equiv \frac{1}{4 C_M \|u_{k}\|_{W^{m+1,p}}}$ and $C_M>0$ is the constant from Morrey's inequality, which only depends on $n,p,\Omega$. Then 
\begin{eqnarray} 
\|\overline{u_{k+1}}\|_{W^{m+1,p}} & \leq &  C_e C_u(k) \Big( \epsilon\: \|\overline{u_{k}}\|_{W^{m+1,p}} + \|\overline{a_{k+1}}\|_{W^{m,p}} \Big), \label{nonlin_estimate_u}  \\
\|\overline{a_{k+1}}\|_{W^{m,p}}  & \leq & \epsilon\: C_e C_a(k) \,  \|\overline{u_{k}}\|_{W^{m+1,p}} , \label{nonlin_estimate_a}
\end{eqnarray} 
where $C_e >0$ is the constant resulting from applying \eqref{Gaffney} and \eqref{Poissonelliptic_estimate_Lp} and 
\begin{align} 
C_u(k) &\equiv  C(M) \big( 1 +  \| u_k \|_{W^{m+1,p}} + \| {u_{k-1}}\|_{W^{m+1,p}} + \| a_{k+1} \|_{W^{m,p}}  \big), \label{diff_Cu}\\
C_a(k) & \equiv  C(M) \big( 1 +  \| u_k \|_{W^{m+1,p}} + \|  {u_{k-1}}\|_{W^{m+1,p}} \big), \label{diff_Ca}
\end{align}
for some constant $C(M) >0$ only depending on $m,n,p,\Omega$ and $M$.
\end{Lemma}

We now establish the induction hypothesis controlling the growth of the iterates allowed by \eqref{nonlin_estimate_u} - \eqref{nonlin_estimate_a} due to nonlinearities, and bound the iterates in the appropriate Sobolev spaces:

\begin{Lemma} \label{Lemma_induction_consistency}              
For some $k\in \mathbb{N}$, assume the induction hypothesis 
\beq   \label{hypothesis_induction}
\|u_k\|_{W^{m+1,p}(\Omega)} \leq 4\, C_0 C_e^2.
\eeq
Let $C_e>1$ and set $\epsilon_1 \equiv  \min\Big( \tfrac{1}{4C^2_e C(M)(1+ 2C_e C_0 + 4C_e^2 C_0)},\tfrac{1}{16 C_M C_0 C_e^2}  \Big)$. 
If $\epsilon \leq  \epsilon_1,$ then, for each $l \in \mathbb{N}$, we have $\epsilon_1 \leq  \epsilon(k+l)$ and the iterates satisfy $\|a_{k+l}\|_{W^{m,p}} \leq  2 C_0C_e$ and $\|u_{k+l} \|_{W^{m+1,p}} \leq  4 C_0 C_e^2$.
\end{Lemma}

Combining the induction assumption \eqref{hypothesis_induction} with our estimates \eqref{nonlin_estimate_u} - \eqref{nonlin_estimate_a} to control the nonlinearities, we prove in \cite{ReintjesTemple_ell2} that for $\epsilon \leq \epsilon_1$ the estimate  
\begin{eqnarray}
\|\overline{u_{k+1}}\|_{W^{m+1,p}}   + \|\overline{a_{k+1}}\|_{W^{m,p}} 
&\leq & \epsilon\: C(M) \:\|\overline{u_{k}}\|_{W^{m+1,p}} , \label{decay} 
\end{eqnarray} 
holds for some constant $C(M)>0$ which depends only on $m$, $n$, $p$, $\Omega$ and $M$. In the final step, assuming $\epsilon \leq \min(\epsilon_1, \frac{1}{C})$, we use a geometric sequence argument to show that $(u_k,a_k)_{k\in \mathbb{N}}$ is a Cauchy sequence in the Banach space $W^{m+1,p}(\Omega)\times W^{m,p}(\Omega)$ which then implies convergence to a solution to $(u,a)$ of \eqref{pde}.\footnote{Note, convergence of $\psi^*_k$ and $y^*_k$ follows directly from the convergence of $a_{k}$ and $u_k$,  because the auxiliary iterates $\psi^*_k$ and $y^*_k$ are only coupled to the equations for $a_k$ and $u_k$ through the boundary data in \eqref{iteration_eqn_J}, which we estimate using the ``Trace Theorem'' together with elliptic estimates and bounds on the nonlinear sources in terms of $u_k$, c.f. \cite{ReintjesTemple_ell2}.}    

We now prove the curvature bound \eqref{bound_u_a_Thm}. By Lemma \ref{Lemma_induction_consistency}, using also the convergence of $u_k$ to $u$ in $W^{m+1,k}$ and convergence of $a_k$ to $a$ in  $W^{m,k}$, we find
\begin{eqnarray}   \label{bound_u_a_eqn1}
\|u\|_{W^{m+1,p}} \leq 4\, C_0 C_e^2 
\hspace{1cm} \text{and} \hspace{1cm}
\|a\|_{W^{m,p}} \leq  2 C_0C_e.
\end{eqnarray}
Now, by our scaling ansatz \eqref{ansatz_scaling} we have $J=I + \epsilon J^*$, $\Gammati = \epsilon\Gammati^*$ and $A=\epsilon a$, and thereby $(I-J,\Gammati) = \epsilon u$. So, using that $\epsilon C_0 = \|(\Gamma, d\Gamma)\|_{W^{m,p}(\Omega)}$ by \eqref{small_Gamma} and \eqref{Gamma-bound}, the bounds in \eqref{bound_u_a_eqn1} imply the sought after estimate \eqref{bound_u_a_Thm}. This completes the proof of Theorem \ref{ThmMain}, (see \cite{ReintjesTemple_ell2} for details). \hfill $\Box$ \vspace{.2cm}

Note, the iteration scheme converges without the need to restrict to a subsequence, and thereby supplies an explicit numerical algorithm for constructing coordinate systems of optimal regularity.

\section{Proof of Theorem \ref{ThmMainCor}} \label{Sec_bounds}

The optimal regularity result of Theorem \ref{ThmMainCor} is a direct consequence of Theorems \ref{Thm_main} and \ref{ThmMain}. Only estimate \eqref{curvature_estimate_Main} of Theorem \ref{ThmMainCor} requires a proof at this point. For this consider the equation
\beq \label{Gammati'_estimate_eqn1}
\Delta \tilde{\Gamma}'  = \Delta \Gammati   -  d \Big( \langle d J^{-1} ;  dJ \rangle  + J^{-1}\langle d J ; \Gamma - \tilde{\Gamma} \rangle \Big),
\eeq
which is derived in \cite{ReintjesTemple_ell1}, c.f.  equation (4.32) of \cite{ReintjesTemple_ell1}.  (Note, equation \eqref{Gammati'_estimate_eqn1} yields $\Delta \Gammati' \in W^{m-1,p}(\Omega)$ in the proof of Theorem \ref{Thm_main}).  We denote with $\Omega$ the domain in which a solution $(J,\Gammati,A)$ exist by Theorem \ref{ThmMainCor}.  Combining now standard elliptic estimates for the Euclidean Laplacian in \eqref{Gammati'_estimate_eqn1} with estimates for the non-linear right hand side of \eqref{Gammati'_estimate_eqn1}, employing Morrey's inequality and the resulting closedness of $W^{m,p}$ under multiplication, one obtains the estimate 
\small
\begin{align}  \label{Gammati'_estimate_eqn2}
\| \Gammati' &\|_{W^{m+1,p}(\Omega')}    \leq 
\ C(M) \Big( \| \Gammati \|_{W^{m+1,p}(\Omega)} + \|I-J^{-1}\|_{W^{m+1,p}(\Omega)} \|I-J\|_{W^{m+1,p}(\Omega)}  \cr
& \hspace{.7cm} +   \|J^{-1}\|_{W^{m+1,p}(\Omega)} \|I-J\|_{W^{m+1,p}(\Omega)}\big( \|\Gamma \|_{W^{m,p}(\Omega)} + \|\Gammati \|_{W^{m,p}(\Omega)} \big) \Big), 
\end{align}
\normalsize
for every compactly contained subset $\Omega'$ of $\Omega$, where here and subsequently $C(M) >0$ denotes a universal constant depending only on $\Omega', \Omega, p, n, m$ and the initial bound $M$. Now, by Lemma 6.1 in \cite{ReintjesTemple_ell2}, we have  
$$
\|I-J^{-1}\|_{W^{m+1,p}(\Omega)} \leq C(M) \|I-J\|_{W^{m+1,p}(\Omega)}.
$$ 
Recalling that $\epsilon\, C_0 = \|(\Gamma, d\Gamma)\|_{W^{m,p}(\Omega)}$ by assumption \eqref{Gamma-bound}, estimate \eqref{bound_u_a_Thm} of Theorem \ref{ThmMain} thus gives us 
\begin{eqnarray} \label{Gammati'_estimate_eqn3} 
\| \Gammati \|_{W^{m+1,p}(\Omega)} + \|I-J\|_{W^{m+1,p}(\Omega)} + \|I-J^{-1}\|_{W^{m+1,p}(\Omega)} \leq  \epsilon\, C(M) C_0  .
\end{eqnarray}
From \eqref{Gammati'_estimate_eqn3}, since $\epsilon >0$ meets the upper bound $\epsilon \leq \epsilon_1$ of Lemma \ref{Lemma_induction_consistency},  we directly obtain the bounds
\begin{eqnarray} \label{Gammati'_estimate_eqn3b}
\|J^{-1}\|_{W^{m+1,p}(\Omega)} \leq  C(M) 
 \hspace{.5cm} \text{and} \hspace{.5cm}
\|J\|_{W^{m+1,p}(\Omega)} \leq  C(M).
\end{eqnarray}
Now substituting \eqref{Gammati'_estimate_eqn3}, \eqref{Gammati'_estimate_eqn3b} and  $\epsilon\, C_0 = \|(\Gamma, d\Gamma)\|_{W^{m,p}}$ into \eqref{Gammati'_estimate_eqn2}, we obtain the estimate
\beq  \label{Gammati'_estimate_eqn4}
\| \Gammati' \|_{W^{m+1,p}(\Omega')}  \ \leq\  C(M) \, \|(\Gamma, d\Gamma)\|_{W^{m,p}(\Omega)}.
\eeq
By \eqref{def_tensor_Gammati}, the connection in $y$-coordinates is $(\Gamma_y)^\alpha_{\beta\gamma} = J^\alpha_k (J^{-1})^i_\beta (J^{-1})^j_\gamma \, \Gammati'^k_{ij}$, so
\beq \label{Gammati'_estimate_eqn5}
\| \Gamma_y \|_{W^{m+1,p}(\Omega')} \leq C_M \|\Gammati'\|_{W^{m+1,p}(\Omega')} \| J^{-1}\|^2_{W^{m+1,p}(\Omega')} \| J\|_{W^{m+1,p}(\Omega')},
\eeq
by Morrey's inequality \eqref{Morrey_textbook}.  Estimating the right hand side in \eqref{Gammati'_estimate_eqn5} by \eqref{Gammati'_estimate_eqn4} and \eqref{Gammati'_estimate_eqn3b}  gives the sought after estimate \eqref{curvature_estimate_Main} and completes the proof of Theorem \ref{ThmMainCor}. \hfill $\Box$

\section{Discussion of the Cauchy problem in General Relativity}     \label{Sec_Discussion} 

We now discuss the difficulties one encounters when trying to smooth non-optimal connections to optimal regularity using hyperbolic PDE methods in the $3+1$ formulation of the Cauchy problem in General Relativity. The $3+1$ framework derives regularity of a solution from the regularity of Cauchy data by PDE methods, assuming a gauge condition \cite{Choquet}. The gauge condition determines a coordinate system in which the regularity of the spacetime metric can be measured.  The current $3+1$ hyperbolic PDE methods for the Einstein equations require the assumption that the induced metric be one derivative more regular than the second fundamental form, (or alternatively that the second fundamental form must be one order more regular than the curvature, \cite{Klainermann}), and deduce from this that the spacetime metric has the regularity of the induced metric given on the Cauchy surface \cite{Choquet}. The second fundamental form accounts for the embedding of the induced metric, and correspondingly its formula involves the connection coefficients from the ambient spacetime, so the second fundamental form, in general, inherits the regularity of the spacetime connection.  Thus to use the $3+1$ framework to regularize a non-optimal metric by one order (when its connection has the same regularity as the curvature), one has to find a gauge condition and a Cauchy surface such that the induced metric and induced second fundamental form both have one more order of regularity than they exhibited in the original non-optimal spacetime coordinate system. The difficulty is that although the induced metric is positive definite, and might be regularized using harmonic coordinates for that metric,  the fact that the formula for the second fundamental form involves the spacetime connection, means the problem of regularizing the second fundamental form on a Cauchy hypersurface in a coordinate gauge that also regularizes the metric on that surface appears formidable. Moreover, all of this has to be accomplished while coupling the hyperbolic equations to the matter model, and hyperbolic PDE's at low regularities are difficult, (e.g., the deep analysis in \cite{Klainermann} required a thousand pages to complete just the vacuum case).\footnote{Referees have requested that we address the issue as to whether the recent resolution of the $L^2$-boundedness conjecture in vacuum spacetimes \cite{Klainermann} has implications to the problem of optimal regularity. First, the $L^2$ theory in \cite{Klainermann} neither addresses nor identifies the problem of optimal regularity. The $L^2$ theorem as stated in Theorem 1.6 of \cite{Klainermann} assumes a weak solution of the vacuum Einstein equations, together with assumptions regarding the restriction of the solution to a Cauchy surface, and from this deduces regularity of the spacetime curvature.  So Theorem 1.6 of \cite{Klainermann}, being based on the $3+1$ formulation of the Cauchy problem in GR, requires the assumption that the second fundamental form be one order more regular than the curvature on that Cauchy surface, which is not true for non-optimal solutions.  Finally, the $L^2$ theorem applies currently only to vacuum spacetimes, while our results apply to general $L^\infty$ connections on tangent bundles of arbitrary manifolds, regardless of matter sources, symmetries or metric signature, a setting even more general than General Relativity.}

We conclude that the $3+1$ framework will estimate a non-optimal solution as being one order less regular than it really is, unless a procedure is given for finding a gauge condition and a Cauchy surface such that the induced metric and induced second fundamental form both have one more order of regularity than they exhibit in the original non-optimal spacetime coordinate system.  Without such a procedure, the Cauchy problem estimates non-optimal solutions as one order less regular than they really are, and in this sense, the Cauchy problem is incomplete in each Sobolev space. Uhlenbeck's methods in \cite{Uhlenbeck} are based on elliptic estimates derived from a Laplace-Beltrami type operator of a positive definite metric in Coulomb gauge. Making the same ideas work to obtain the results in \cite{Uhlenbeck} using the wave operator associated with the Laplace-Beltrami operator of Lorentzian metrics in harmonic coordinates, is problematic because of the need to regularize initial data required for the evolution.                 

In summary, our results are independent from, do not compete with, and complement the body of literature on the GR Cauchy problem because we establish that non-optimal solutions can always be regularized to meet the assumptions required for the second fundamental form in $3+1$ formulations of the initial value problem.    In our theory optimal regularity is obtained by coordinate transformations constructible by explicit algorithms for generating solutions of the elliptic RT-equations, for general $L^\infty$ connections on the tangent bundle of arbitrary manifolds, regardless of matter sources, regardless of symmetries, and regardless of metric signature.   This does not follow from the current technology of the Cauchy problem in GR even in the case of vacuum, essentially because those methods derive regularity of solutions from the induced metric (or curvature) and induced second fundamental form on Cauchy surfaces, starting with the assumption that the induced metric is one derivative more regular than the second fundamental form, and for non-optimal connections, they are both one order below optimal.

\section*{Concluding remarks}

A major problem in Mathematical Physics is how to extend the estimates obtained by elliptic methods in Riemannian geometry, (e.g. Uhlenbeck's papers \cite{Uhlenbeck,Uhlenbeck_first}, topic of the 2019 Abel Prize and 2007 Steele Prize), to the Lorentzian setting of Physics. Uhlenbeck's results, (Theorems 1.3 and 1.5 of \cite{Uhlenbeck}), use the elliptic Laplace-Beltrami operator of an underlying Riemannian metric to estimate the associated connection as one derivative above the curvature, by eliminating uncontrolled derivatives in the connection via transformation to Coulomb gauge.  The extra derivative then yields Uhlenbeck's celebrated compactness result. A dominant point of view seems to be that to extend the Riemannian methods of \cite{DeTurckKazdan,Uhlenbeck} to the Lorentzian case, one must derive results from the non-linear wave equation in harmonic type coordinates, analogous to these elliptic methods of Riemannian geometry.   However, the RT-equations introduced here show that, associated with any connection on the tangent bundle of an arbitrary manifold, there exists an elliptic system entirely different from the Laplace-Beltrami based elliptic system,  and this system also lifts the regularity of the connection one order above the curvature--but it applies to metric connections associated with Riemannian, Lorentzian and semi-Riemannian metrics alike, independent of metric signature.  Our take on this is that the hyperbolic PDE approach is too complicated, and the elliptic RT-equations are simpler, essentially because extracting regularity from initial data in a hyperbolic problem is entirely different from extracting regularity from source terms in elliptic problems.\footnote{See \cite{Nardmann} for the example of the prescribed scalar curvature problem, in which the hyperbolic approach  also appear not feasible, and construction of an auxiliary Riemannian metric is central for extending results to Semi-Riemannian smooth manifolds. (The results and methods in \cite{Nardmann} are not further related to ours.)}  We propose the RT-equations to bridge the gap in analysis between Riemannian and Lorentzian geometry, by extending elliptic regularity theory to semi-Riemannian manifolds without requiring the assumption of positive-definiteness.

\section*{Acknowledgements}

The authors thank Jos\'e Nat\'ario and the Instituto Superior T\'ecnico for supporting this research. We thank Craig Evans for suggesting reference \cite{Dac}, and for helpful comments on elliptic PDE theory. We thank John Hunter, Pedro Gir\~ao, Steve Shkoller and Kevin Luli for helpful discussions. We thank referees at RSPA for pointing us to Uhlenbeck's paper \cite{Uhlenbeck} and to the doctoral thesis \cite{Nardmann}.

\providecommand{\bysame}{\leavevmode\hbox to3em{\hrulefill}\thinspace}
\providecommand{\MR}{\relax\ifhmode\unskip\space\fi MR }

\providecommand{\MRhref}[2]{  \href{http://www.ams.org/mathscinet-getitem?mr=#1}{#2} }
\providecommand{\href}[2]{#2}

\end{document}